\documentclass{ametsocv6.1}
\usepackage[a4paper]{geometry}
\usepackage{booktabs}
\usepackage{multirow}
\usepackage{tabularx}
\usepackage{subcaption}
\usepackage{caption}
\usepackage{makecell}
\usepackage{url}
\usepackage{fancyvrb}
\usepackage{lineno}
\usepackage[colorlinks=False, hidelinks]{hyperref} 
\usepackage[inline,shortlabels]{enumitem}
\usepackage{graphicx, wrapfig}

\usepackage{xcolor}

\newcommand{\Pf}{p_f} 
\newcommand{\Po}{p_o} 
\usepackage{listings}
\definecolor{yamlgreen}{rgb}{0.0, 0.5, 0.0}
\lstdefinestyle{yaml}{
  basicstyle=\color{yamlgreen}\footnotesize\ttfamily,
  columns=fullflexible,
  showstringspaces=false,
  rulecolor=\color{black},
  string=[s]{'}{'},
  stringstyle=\color{blue},
  comment=[l]{:},
  moredelim=[is][commentstyle]{||}{££}, 
  commentstyle=\color{black},
  morecomment=[l]{-}
  morecomment=[s]{]}
  frame=none,
}

\nolinenumbers

\title{Observation-guided Interpolation Using Graph Neural Networks for High-Resolution Nowcasting in Switzerland}
\authors{Oph\'elia Miralles\aff{a,b}\correspondingauthor{Oph\'elia Miralles, ophelia.miralles@usys.ethz.ch}, Daniele Nerini\aff{b}, Jonas Bhend\aff{b}, Baudouin Raoult\aff{c}, Christoph Spirig\aff{b}
}

\affiliation{\nolinenumbers \aff{a}{Center for Climate Systems Modeling (C2SM), EHTZ, Z\"urich} \\
	\aff{b}{MeteoSwiss, Z\"urich}\\
	\aff{c}{European Centre for Medium-Range Weather Forecasts (ECMWF), Reading, UK}}

\abstract{Recent advances in neural weather forecasting have shown significant potential for accurate short-term forecasts. However, adapting such gridded approaches to smaller, topographically complex regions like Switzerland introduces computational challenges, especially when aiming for high spatial (1 km) and temporal (10 minutes) resolution.  This paper presents a Graph Neural Network (GNN)-based approach for high-resolution nowcasting in Switzerland using the Anemoi framework and observational inputs. The proposed architecture combines surface observations with selected past and future numerical weather prediction (NWP) states, enabling an observation-guided interpolation strategy that enhances short-term accuracy while preserving physical consistency. We evaluate two models, one trained using local nowcasting analyses and one trained without, on multiple surface variables and compare it against operational high-resolution NWP (ICON--CH1) and nowcasting (INCA) baselines. 
Results over the test period show that both GNNs consistently outperform ICON--CH1 when verified against INCA analyses across most variables and lead times. Relative to the INCA forecast system, scores against INCA analyses show AI gains beyond 2h (with early--lead disadvantages attributable to INCA’s warm start from the analysis), while verification against held--out stations shows no systematic degradation at short lead-times for AI models and frequent outperformance across surface variables. A comprehensive verification procedure, including spatial skill scores for precipitation, pairwise significance testing and event-based evaluation, demonstrates the operational relevance of the approach for mountainous domains. 
These results indicate that high--resolution, observation--guided GNNs can match or exceed the skill of established forecasting systems for short lead times, including when they are trained without nowcasting analyses.}

\begin{document}

\maketitle
\statement
The World Meteorological Organization (WMO) defines nowcasting as forecasting with local detail, by any method, over a period from present to six hours ahead, including a detailed description of the present weather. Nowcasting supports weather-sensitive decisions and typically focuses on near-surface variables. These fields are strongly shaped by complex topography, which induces local flow patterns that are difficult to predict. We develop a deep-learning model that fuses local topographic information with radar, satellite, station, and NWP inputs to produce short-term forecasts comparable to those from MeteoSwiss' operational nowcasting system (INCA).

We address two questions: 
\begin{enumerate*}[i)]
    \item can a model trained to emulate nowcasting analyses match operational nowcasting forecast quality, and
    \item can a model trained without any nowcasting analyses, but using only numerical weather prediction forecasts, topography, and observations, achieve comparable skill?
\end{enumerate*}
The second question is crucial because many meteorological services lack nowcasting analyses for all near-surface variables; a method that attains INCA-like performance without such analyses would therefore have broad operational value.

\clearpage
\section*{Introduction}
Nowcasting, also referred to as very short-term forecasting, involves rapid correction of the most recently available numerical weather prediction (NWP) model run with real-time observations. Traditional nowcasting systems such as the MeteoSwiss operational nowcasting system INCA (Integrated Nowcasting through Comprehensive Analysis), combine three main approaches \citep{haiden2011integrated}: Lagrangian extrapolation of analysis fields when available (for example, for cloudiness or precipitation), interpolation of residuals at sparse locations with surface measurements and simple blending with NWP model outputs. The nowcasted variable of interest is typically a weighted sum of these three terms, in which the weight attributed to NWP model output is low for short lead times as persistence of observations is more skillful, and increases with the lead time in order to favour seamlessness. The nowcasting lead time range is typically from zero to six hours with a sub-hourly time granularity corresponding to the observation update rate. The spatial resolution for nowcasting output is equally high, usually on the order of 1km. Operational nowcasting is mostly used for weather-dependent decision-making and is thus based on surface variables. In contrast, NWP forecasts a wide range of variables across multiple pressure levels up to ten days ahead with coarser temporal granularity. Nowcasting requires fully automated, ultra-fast data ingestion and processing to produce reliable outputs in very short lead times (0 to 6 h), often updating every few minutes.

Recent work in machine learning, particularly deep learning, has focused on improving precipitation nowcasting from radar data \citep{shi_convolutional_2015,shi_deep_2017,agrawal2019machine,leinonen2020stochastic,ravuri2021skilful, zhang2023skilful}, with notable success in capturing extreme \citep{zhang2023skilful} or convective \citep{ravuri2021skilful} events. 
Deep learning models offer a fundamental shift from conventional nowcasting approaches by learning the nonlinear development of precipitation fields, rather than relying on simplistic assumptions of persistence and advection. This enables them to anticipate complex spatio-temporal patterns such as rapid intensification and localized extremes. Those characteristics, which cause the greatest damage and socioeconomical impacts \citep{ravuri2021skilful, zhang2023skilful}, are typically smoothed out or missed by traditional methods. Importantly, our work highlights that leveraging the joint correlations of surface variables (beyond precipitation alone) can provide richer predictive signals, an aspect that remains underexplored in most existing deep learning models focused only on precipitation.

Observation-driven methods aiming to forecast several surface variables at the same time show encouraging first results on the nowcasting range in the United States of America \citep{sonderby2020metnet,andrychowicz2023deep}. The MetNet3 approach, for example, combines a U-Net architecture with a visual transformer to assimilate diverse weather data and generate reliable forecasts on large domains \citep{andrychowicz2023deep}. Although the contributions of the MetNet team are influential \citep{sonderby2020metnet, andrychowicz2023deep}, the reproducibility of their models remains challenging due to limited public documentation and implementation details.

Most existing approaches use U-Net–like architectures tailored to large, relatively flat domains. In topographically complex regions like Switzerland, operational nowcasting faces major challenges, including terrain-modulated wind and temperature fields, sharp spatial gradients, and the need for seamless integration with NWP. Building on recent successful data-driven regional modelling efforts using Graph Neural Networks (GNNs) \citep{price2024gencast, lang2024aifscrps, lang2024aifs, alexe2024graphdop, nipen2024regional}, we propose an alternative approach inspired by GraphCast \citep{lam2022graphcast} to nowcast wind, temperature, precipitation and humidity in Switzerland. GNNs enable flexible spatial modeling and efficient computation over irregular domains. Their structure is also well suited to integrating diverse data sources with heterogeneous spatial supports, such as radar, stations, and satellite observations, making them a promising framework for future extensions.

In the graph neural network (GNN) described in this paper, atmospheric interactions are modelled as a graph, with each geographical location (such as a grid cell or weather station) represented by a node. The edges between nodes capture the interactions that illustrate how weather patterns at different locations influence each other. The network processes data through multiple hidden nodes, refining the information at various levels of granularity, before mapping the processed data back to specific locations via the output nodes.

For operational nowcasting, it is crucial that the models are transparent and reproducible. Therefore, the study presented here is mainly based on \hyperlink{https://anemoi.readthedocs.io/en/latest/index.html}{anemoi}, an open-source and well-documented Python framework recently introduced and used operationally by the European Centre for Medium-range Weather Forecast (ECMWF).
Anemoi is intended to facilitate the development, integration, and operational inference of deep learning models for real-time weather forecasting.
It comprises training dataset preparation, e.g., common tools to create and pre-process training datasets, configurable, command-line-based model training and inference, and a catalog of weather datasets and trained models open to its community and end users.

This study contributes to operational nowcasting by introducing a multi-variable deep learning approach that leverages diverse surface observations. The specific goals of this study include: \begin{enumerate*}[i)]
    \item emulate INCA analyses at 1.1 km resolution using a GNN for low-latency forecasts,
    \item evaluate whether AI-based forecasts can match or exceed the skill of INCA and ICON--CH1over complex terrain, and
    \item assess the potential of replacing nowcasting analyses with data-driven AI models trained only on topography, observations and NWP.
\end{enumerate*}

It is organised as follows. Section~\ref{sec:data} describes the data used for the experiments of Section~\ref{sec:experiments}. The specific samples used for training are further detailed in Section~\ref{sec:data_dl}. The technical details of the training are mentioned in Section~\ref{exp1}.
The verification procedure is then presented in Section~\ref{sec:verif}, followed by a detailed analysis of the results. The paper concludes with Section~\ref{sec:conclusion} which offers insight and explores future challenges and potential directions.

\section{Description of Data Sources}
\label{sec:data}
This section describes each data source without detailing their role in the deep learning model, which is expanded in further detail in Section~\ref{sec:data_dl}.

\subsection{Observations}
\subsubsection{Station data}
\label{subsec:station}
\begin{figure}
    \centering
\includegraphics[width=.6\linewidth,]{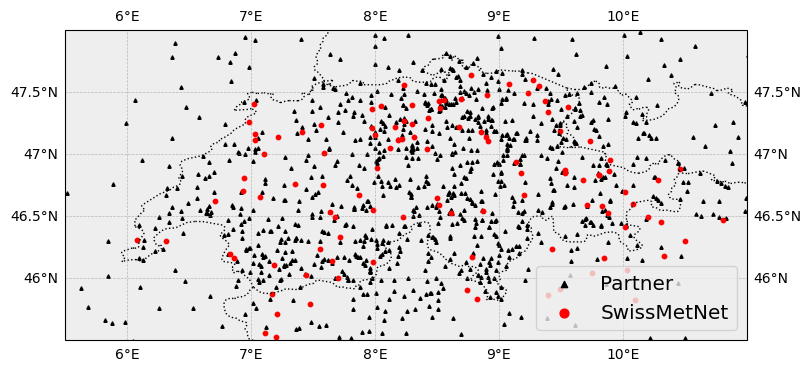}
\includegraphics[width=.39\linewidth, height=.2\textheight]{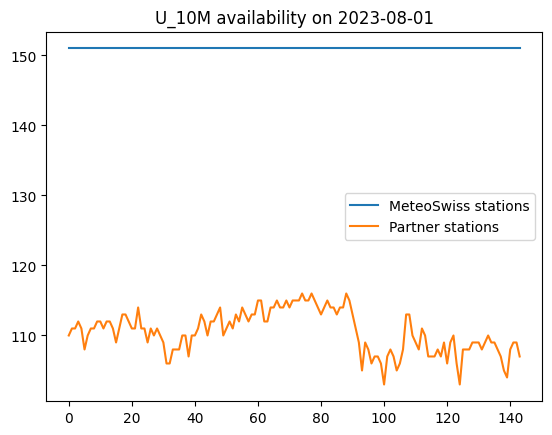}
\caption{Left: all available weather stations from SwissMetNet (red bullets) and partner (black triangles) stations. Number of MeteoSwiss/partner stations available per period of 10 minutes on Aug 1 2023 (right).}
\label{fig:station}
\end{figure}

Station observations were sourced from the MeteoSwiss Data Warehouse (DWH). Due to the highly variable availability of partner stations over time (as shown on a single date in Figure~\ref{fig:station}), we selected the 151 SwissMetNet stations with nearly no missing values during the training period. The leftover (partner) stations are used to measure the performance out-of-sample in the verification.

\subsubsection{Radar}
In addition to station data, radar-derived composite data is used for precipitation analysis. The radar product PRECIP \citep{peleg_twelve_2019} provides ground-level precipitation intensity estimates based on a weighted aggregation of reflectivity from all radars above a given pixel. This data is corrected for visibility, reflectivity, and global and local biases. Quality control also involves removing observed rain rate that is greater than 150mm per hour. The radar composite covers only parts of the Swiss radar domain (Figure~\ref{fig:rad_sat}). Finally, although the proprietary MeteoSwiss radar data is not open-access, a composite rain rate product covering Switzerland is available via the OPERA API (\href{https://donneespubliques.meteofrance.fr/}{donneespubliques.meteofrance.fr}) for operational and research purposes. 

\subsubsection{Satellite}
\begin{figure}
    \centering
\includegraphics[width=.49\linewidth]{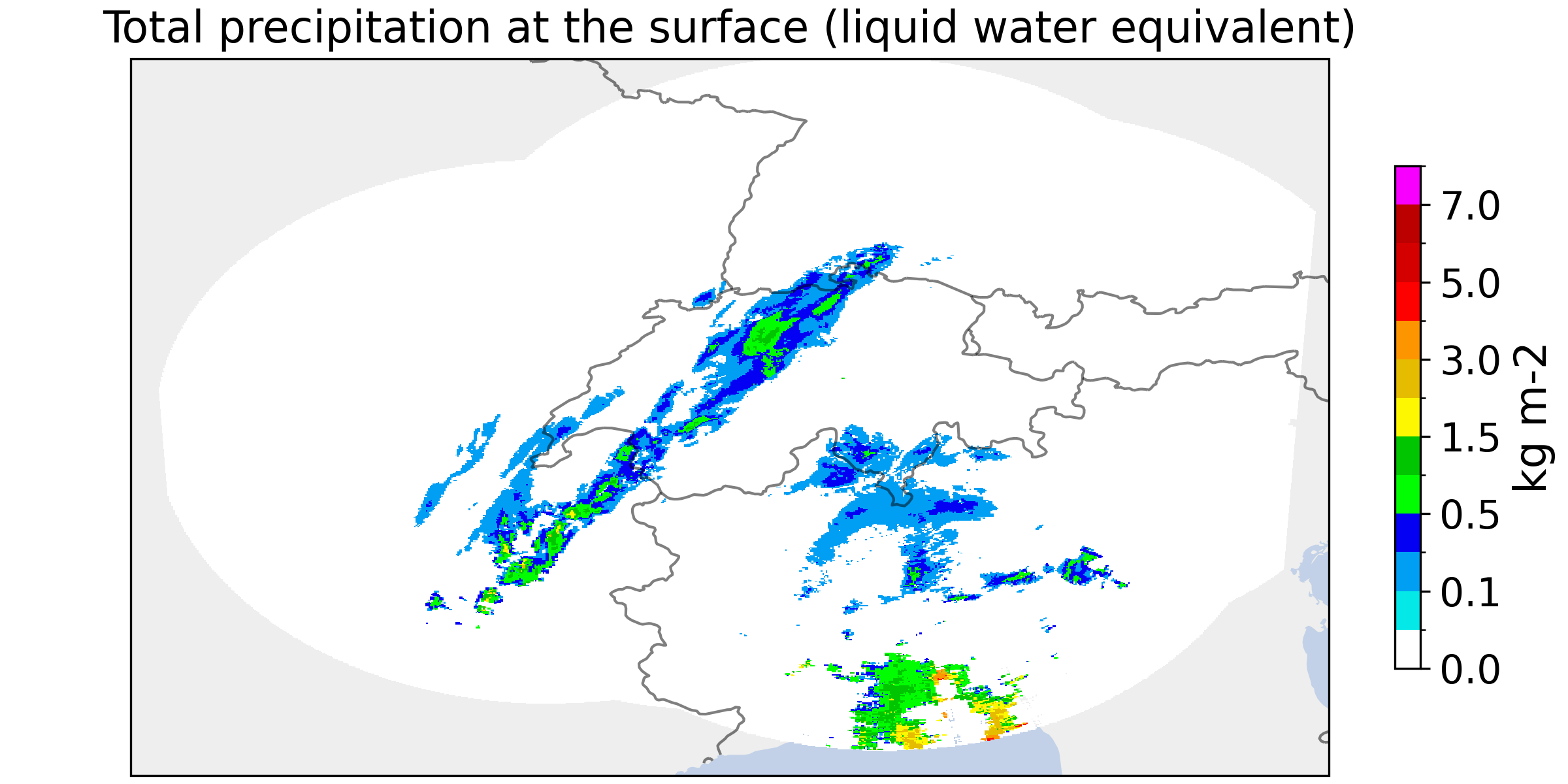}
\includegraphics[width=.49\linewidth]{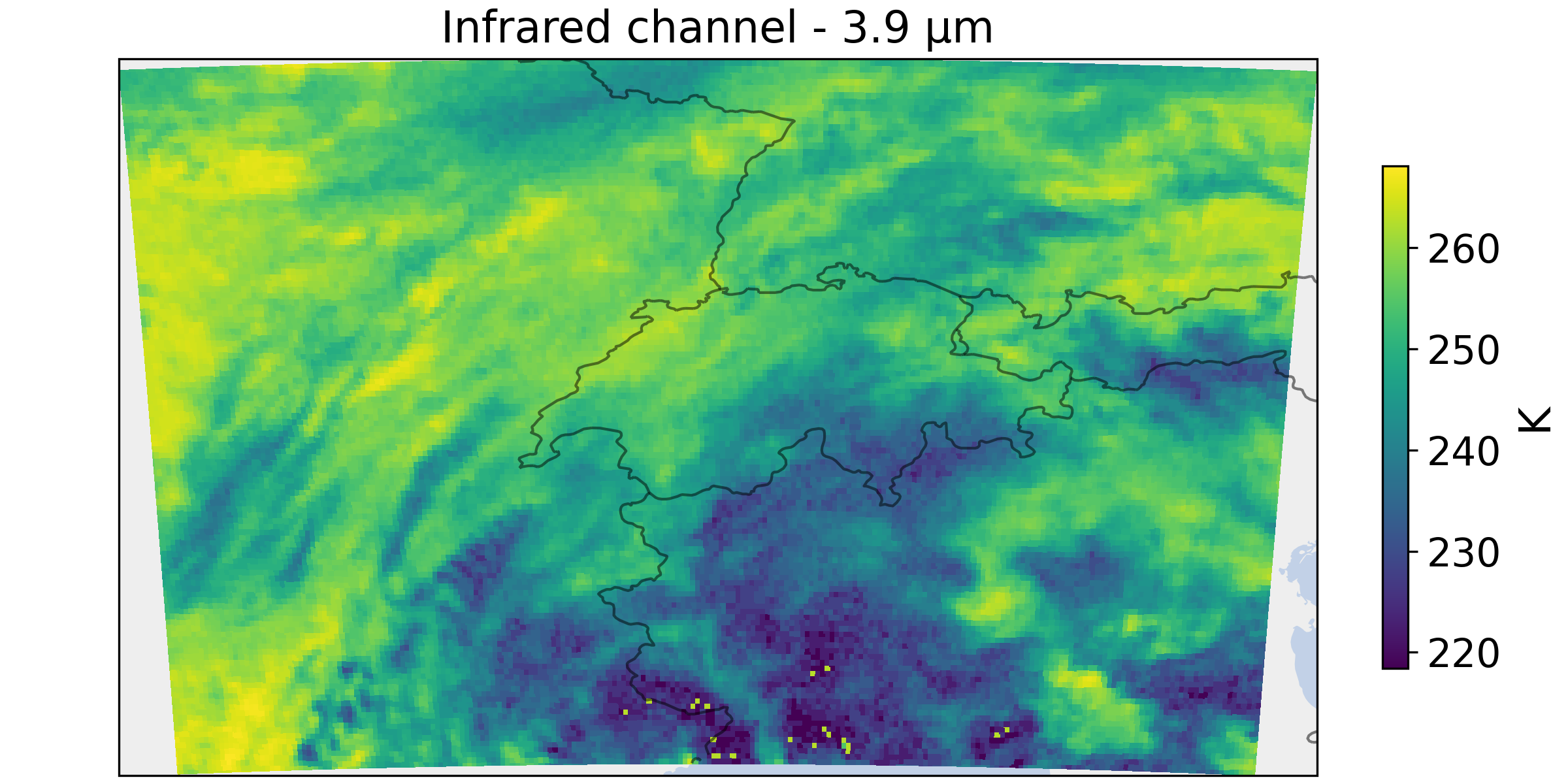}
\includegraphics[width=.49\linewidth]{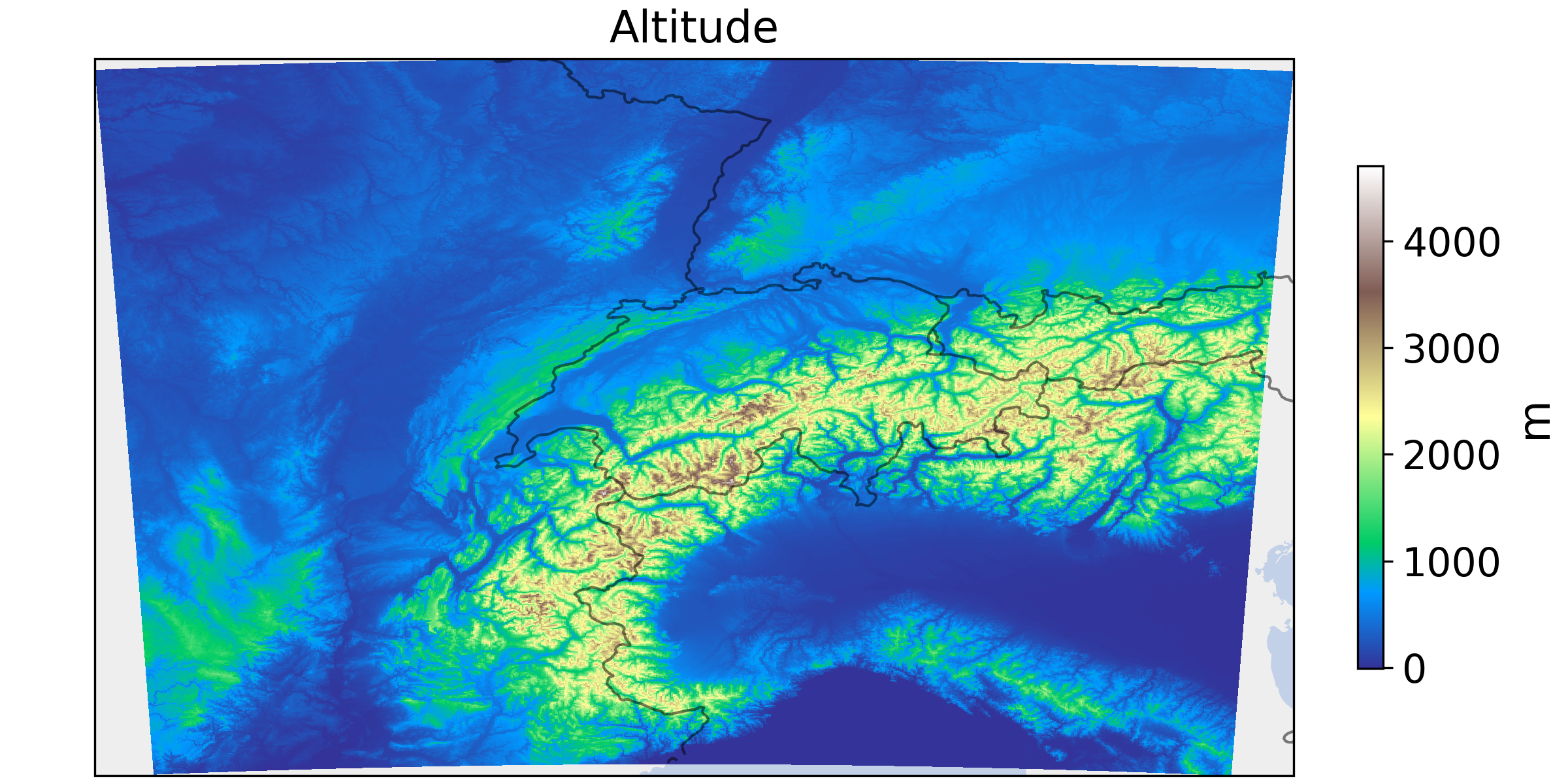}
\caption{Composite and raster data: rain rate from radar (top left), infrared channel from MSG satellite (top right), topography (bottom).}
\label{fig:rad_sat}
\end{figure}

Satellite raw infrared and visible channels are also used as input to the network (see, e.g., Figure~\ref{fig:rad_sat}). We chose to use the channels presented in Table~\ref{tab:sat} based on expert judgment of their relevance for nowcasting surface variables.
\begin{table}[h]
\centering
\begin{tabular}{|l|p{12cm}|}
\hline
\textbf{Channel} & \textbf{Specificity} \\
\hline
\texttt{IR\_108} & Has good surface penetration under clear skies. Useful for surface temperature and indirect wind cues. \\
\hline
\texttt{IR\_016} & Sensitive to low clouds, aerosols, and snow/ice. Useful for surface analysis, especially during the day. Helps estimate dewpoint indirectly by identifying surface moisture signatures. \\
\hline
\makecell[l]{\texttt{VIS006} \\ and \texttt{HRV}} & Detects clouds, fog, and surface features. Good for observing low-level cloud movement, which helps infer surface wind and humidity distribution. \\
\hline
\texttt{IR\_039} & Useful for fog and low cloud at night, can help indirectly. \\
\hline
\end{tabular}
\caption{Selected satellite channels and their primary specificities relevant for short-term forecast of surface variables. All channels were first considered, and then narrowed down to the most relevant using expert knowledge.}
\label{tab:sat}
\end{table}
The Meteosat Second Generation (MSG) system consists of two geostationary satellites positioned at $0^{\circ}$E and $9.5^{\circ}$E. These satellites provide images with a spatial resolution of approximately 3 km (east-west) by 5 km (north-south). Every 15 minutes, a full-disk scan captures an image of the entire Earth, while a rapid scan focuses on the upper third of the Earth every five minutes. The satellites alternate their roles, with one conducting the full-disk scan while the other performs the rapid scan. We use full-disk scans from the open-access EUMETSAT API (\href{https://user.eumetsat.int/data-access}{eumetsat.int}) for easy reproducibility.

\subsubsection{Topography}
Switzerland's terrain is complex: local topographic characteristics strongly modify surface wind speeds and temperatures, and to allow the network to learn these relationships, we use the topography of the freely available 90-meter resolution SRTM3 digital elevation model (DEM) constructed by NASA and NGA \citep{jarvis2008}, Figure~\ref{fig:rad_sat}. High resolution topography is retrieved on the Swiss radar grid and the nearest point to each grid cell is selected, thus providing the deep learning model with precise information about elevation.

\subsection{Nowcasting}
\label{sec:incadata}
MeteoSwiss’s INCA nowcasting system updates every 10 min for most variables; precipitation updates every 5 min, synchronized with radar availability. INCA provides continuous forecasts with 1km resolution on the Swiss radar domain. It seamlessly combines observed, extrapolated, and predicted data from the deterministic ICON--CH1run. The current deterministic system runs a single ensemble member and has two flavors: an analysis and a forecast. The analysis refers to an ex post high resolution, observation-informed estimate of the current state of the atmosphere at a specific time ($t_0$), serving as a reference for evaluating short-term forecasts. The forecast is available every 10 minutes for precipitation and hourly for other surface variables and predicts the next 6 hours. In this study, analysis data for 10-meter wind, 2-meter dewpoint, temperature, and rain rate \citep{2020siderisnowprecip} serve as "ground truth" and INCA forecast as a baseline model.

\subsection{Numerical Weather Prediction}
\label{sec:icon}
We use ICON, the ICOsahedral Non-hydrostatic modelling framework originally developed by Deutscher Wetterdienst and MPI \citep{zangl2015span}, as the NWP reference. ICON--CH1data calibrated for Switzerland (ICON--CH1, CH1 standing for Switzerland -CH- at 1km resolution) is available at a spatial resolution of 1km, is updated every 3 hours, and uses ECMWF-IFS as boundary conditions. Although ICON--CH1 is reinitialized every 3 hours and is typically made available to downstream applications such as INCA within 1-2 hours after initialization, in this study we chose a 12-hour update interval. This choice reflects both a conservative assumption regarding the realistic availability of input data within the forecast window and practical considerations during the construction of the Anemoi datasets.

When this work was conducted, ICON--CH1 forecasts were available from August 2023 onward only. The amount of data appears sufficient relative to the size of the model compared to the recent literature \citep{lang2024aifs, nipen2024regional}. While two years of 10-minute data provide a greater number of data points than fifty years of 6-hour slices, we acknowledge that higher temporal resolution does not necessarily translate to proportionally more independent information due to strong temporal and spatial correlations inherent in weather data. However, for nowcasting applications, high-frequency observations are essential as they enable the model to capture rapid changes and short-term dynamics that coarser data may overlook, thereby enhancing immediate forecasts. We also recognize that the dataset may not include sufficient variability or extreme events, particularly those occurring outside the training period, which limits the model’s ability to learn these phenomena.

A wide range of variables were initially considered, including standard surface variables, such as 2-meter temperature and dewpoint, 10-meter wind components and rainfall rate, as well as model-level fields for relative humidity (Q), temperature (T), wind components (U, V) and pressure (P) across the 80 terrain-following vertical model levels. More details on ICON--CH1model levels and their relation to altitude can be found in Section 3.4 of \citet{reinert_icon_2024}. In the end, only surface variables were retained, as upper-level variables did not demonstrate a significant predictive ability for short-term prediction in this study.

\section{General architecture}
\label{sec:experiments}
\subsection{Model}
The input data on the original data mesh is first ``encoded" into a weather message through attention layers and projected onto a smaller set of nodes, called the hidden mesh, where weather messages are aggregated. The graph topology (or node locations) defines which nodes exchange information (nearest neighbours, terrain-aware links), but the edge weights are computed from the current predictors and therefore vary with the flow and regime.

Concretely, during message passing the update at node \(i\) takes the form
\begin{equation}
h_i^{(\ell+1)} \;=\; \phi\!\left(h_i^{(\ell)}, \;\sum_{j\in\mathcal N(i)} \alpha_{ij}(x)\,\psi\!\left(h_i^{(\ell)},h_j^{(\ell)}\right)\right),
    \label{eq:mes-pas}
\end{equation}
where:
\begin{itemize}
  \item \(h_i^{(\ell)}\) is the hidden state at node \(i\) and layer \(\ell\), encoding local predictors;
  \item \(\psi(\cdot)\) is the learnt message passing function that maps the sender/receiver states and optional edge features \(e_{ij}\) (e.g., distance, azimuth, elevation/slope difference) to a message vector;
  \item \(\alpha_{ij}(x_t)\in[0,1]\) are data-dependent (attention-like) weights computed from the inputs \(x\) (e.g., local wind components, stability, orography);
  \item \(\phi(\cdot)\) is the node update that combines the previous layer's state and the aggregated message from surrounding nodes to compute the next layer's state.
\end{itemize}
Thus, when the synoptic flow turns from North-West to South-East, the model reweights messages to favour downwind neighbours; when the flow shifts, the effective connectivity shifts with it. The dependence on the predictors \(x\) is implicit through the hidden node states \(h_i^{(\ell)}\) and edge features \(e_{ij}\); we do not explicitly write the conditioning on \(x\) in \(\psi\) and \(\phi\) in Equation~\ref{eq:mes-pas} for clarity. Instead, \(x\) is explicit in Equation~\ref{eq:mes-pas} only where it directly changes the weights (without intermediary step, e.g., in \(\alpha_{ij}(x)\)).

``Processing" then takes place in the smaller set of nodes, to limit the computational burden. Aggregated weather messages are exchanged throughout neighbouring nodes along the edges, thus accounting for spatial correlation. The resulting graph, which contains information on location specificities through the nodes and spatial correlation through the edges, is finally ``decoded" and projected back to the original grid. This model architecture has been extensively described in the recent literature on weather forecasting using graph neural networks \citep{lam2022graphcast, lang2024aifs, lang2024aifscrps, alexe2024graphdop}, so is not further detailed here. The main distinction from existing medium-range forecasting studies using GNNs lies in the higher temporal resolution, emphasis on surface predictors, and the use of a smaller, higher-resolution spatial domain (with overall more nodes on the graph).

Note that we use autoencoder terminology in the schematic purely for intuition: the model is trained end-to-end as a single network. Likewise, the ``processor" denotes the latent-space computation layers that sit between the encoder and decoder, but these layers are trained jointly with them rather than separately.

\begin{figure}[ht]
\centering
\includegraphics[width=.8\linewidth]{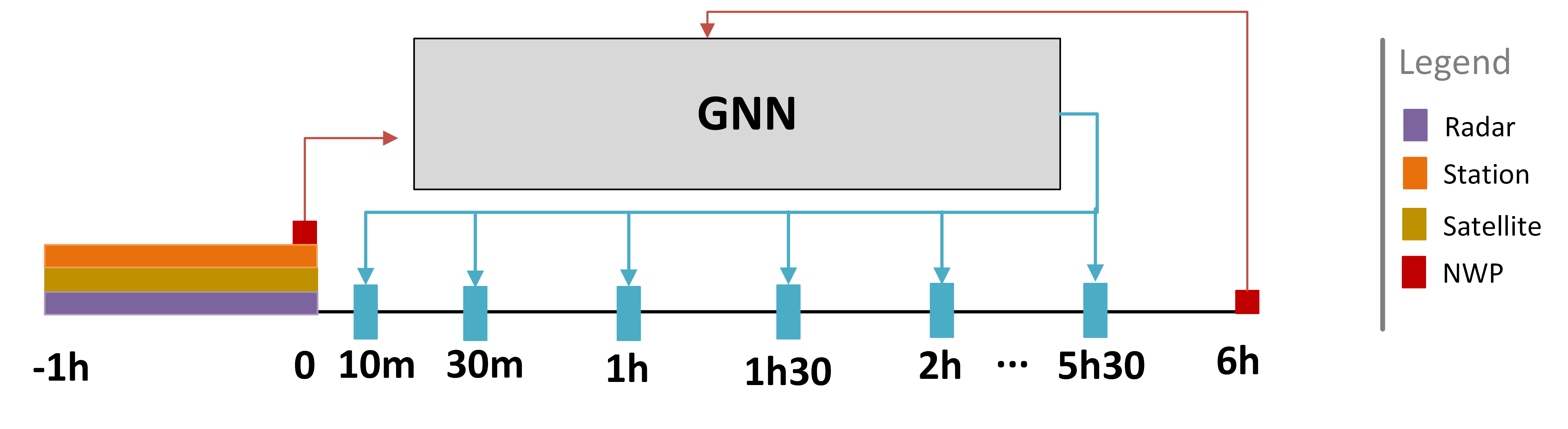}
\caption{Schematic representation of the observation-guided interpolation model used for nowcasting. This is the model used for training; at inference time, we use the last hour of observations and [$t_0, t_0+12$h] from ICON--CH1 to forecast every 10 minutes over the next 12-hours.}
\label{fig:exp1}
\end{figure}

The model uses observations and NWP input to predict surface variables of interest. This approach incorporates both past and future NWP states as input to the GNN, allowing it to perform an observation-guided interpolation between these states (Figure~\ref{fig:exp1}).

\subsection{Loss function}
\label{sec:loss}
In the context of this study, nowcasting analysis data was available over Switzerland. We start by training an initial model, referred to as AINCA (AI-based version of INCA forecasting system), which is evaluated using a pointwise loss against the INCA analysis (Section~\ref{sec:incadata}) used as the ground truth. However, such analysis data may not be readily accessible in all countries, particularly within the nowcasting time frame. To address this limitation, another model, called $\mathcal{L}$-AINCA, explores the forecasting capabilities of Graph Neural Networks (GNNs) in the absence of nowcasting analysis data, necessitating the use of a more sophisticated loss function.

The loss function for the model trained on station data should incorporate two key aspects of nowcasting. First, the influence of observations should gradually decrease as the lead time increases, aligning with the numerical weather prediction (NWP) update when it becomes available. Second, the model should capture the physical patterns present in the NWP data while closely matching the exact station data at the observation locations. This requires the use of different metrics to evaluate the distances from the NWP and the observational data.  

\paragraph{Pointwise loss}
The Huber loss offers a compromise between the mean squared error (MSE) and the mean absolute error (MAE), providing sensitivity to small residuals while being more robust to outliers. This makes it particularly suitable in settings where occasional large errors may otherwise disproportionately influence optimization. We employ the Huber loss for pointwise comparison:
\begin{equation}
\mathcal{L}_{\text{H}}(y, \hat{y}) =
\begin{cases}
\frac{1}{2} (y - \hat{y})^2, & \text{if } |y - \hat{y}| \leq 1, \\
 \left( |y - \hat{y}| - \frac{1}{2} \right) & \text{otherwise},
\end{cases}
\label{loss_exp1}
\end{equation}
where $y$ is the input vector and $\hat{y}$ its estimate. 

\paragraph{Spatial loss}
We use the logarithmic spectral distance (LSD) to assess whether the generated images preserve the spatial structures observed in the target images. The LSD \citep{rabiner1993fundamentals} is the logarithmic difference in power spectra between the generated and realised samples. The node-weighted version can be expressed as
\[\mathcal{L}_{\mathrm{LSD}}\left(y, \hat{y}\right) = \left\Vert 10\log_{10}\left( \frac{\vert f(y)\vert^2}{\vert f\left(\hat{y}\right)\vert^2} \right)\right\Vert_2,\]
where $f$ is the Fourier transform, $\Vert.\Vert_2$ is the $\mathcal{L}_2$-norm and $\vert f(\cdot)\vert^2$ the power spectrum. In \citet{yan_fourier_2024}, using a probabilistically weighted combination of LSD and Fourier correlation loss (FCL) yields promising results. The FCL is expressed as
\[\mathcal{L}_{\mathrm{FCL}}(y, \hat{y}) = 1 - \frac{ \mathrm{Re}\left[f\left(\hat{y}\right)^\mathrm{T} \Bar{f}\left(y\right) \right]}{\Vert f\left(\hat{y}\right)\Vert_2 \Vert f\left(y\right)\Vert_2},\]
where $\Bar{f}$ denotes the complex conjugate of $f$, $x^\mathrm{T}$ is the transpose of $x$ and and $\mathrm{Re}$ is the real part of a complex number.
We then define the spatial loss as
\[\mathcal{L}_{\text{S}}( \hat{y}, y) = \mathbf{1}_{w \leq 0.5} \mathcal{L}_{\mathrm{LSD}}( \hat{y}, y) 
+ \mathbf{1}_{w > 0.5} \mathcal{L}_{\mathrm{FCL}}( \hat{y}, y),\]
where $\mathbf{1}$ represents the Heaviside step function and $w$ is sampled uniformly between 0 and 1.

\paragraph{Total loss}
We denote by $t \in [0,1]$ the lead time fraction corresponding to the ratio of the predicted lead time over the horizon $H$.
The total loss then writes
\begin{equation}
    \mathcal{L}\left(t\right) = w_1 e^{-\alpha (1-t)} \mathcal{L}_{\text{S}}\left(y_{\rm NWP}, \hat{y}\right) + w_2 e^{-\alpha t} \mathcal{L}_{\text{S}}\left(y_{\rm radar}, \hat{y}\right) + w_3 e^{-\alpha t}\mathcal{L}_{\text{H}}\left(y_{\rm station}, \hat{y}\right),
    \label{loss_exp2}
\end{equation} 
where $\alpha$ is the lead time decay factor. Close to the present time (i.e., $t$ is small), the predictions should be close to the observed values, but when $t$ is close to horizon $H$, more importance is given to the loss comparing the predictions to the NWP values. 

\section{Experimental Setup}
\label{sec:data_dl}
This section details the data transformations, train/validation/test splits, and the specific usage for each source introduced in Section~\ref{sec:data}.

\subsection{Temporal Coarsening of NWP Data}
ICON--CH1 inputs were taken from the operational archive, which provides 3h cycles for Switzerland; higher frequency data were not available when the study was conducted. At wall-clock time \(t\), ICON--CH1 provides the latest available cycle with valid times from \(t\) out to \(\sim t{+}33\)h; the subsequent cycles that will be issued at \(t{+}3\) and \(t{+}6\)h are not yet available. 
As explained in Section~\ref{sec:experiments} and illustrated in Figure~\ref{fig:exp1}, the model is trained to interpolate between \([t,\,t{+}6~\text{h}]\). To produce a 6h forecast starting at \(t\), the model needs to generate the intermediate 10~min steps in \([t,\,t{+}6~\text{h}]\) without relying on future updates. To enforce this and provide flexibility to generate 12hours sequences, we freeze the latest available cycle and use coarse data from the same cycle, that is, the ICON--CH1 slices valid at \(t\), \(t{+}6\), and \(t{+}12\)h.
We mirror this frozen-cycle policy in training: the archive is sliced into 12h windows that use only inputs from a single ICON--CH1 cycle, with no mid-window refresh. Thus, neither training nor inference ever uses information from future cycle updates. 
This choice does not prevent the model from running with the most recent forecast cycle available at $t$ at inference time.
During verification, we sample the baselines and our forecasts every 3h during the test period (\(\approx 240\) timestamps), and the ICON--CH1 fields used correspond to the operational 3-hour update cycles.

\subsection{Spatio-temporal Alignment}
\label{sec:temp_agg}
Data from the various sources described in Section~\ref{sec:data} are either aggregated or interpolated to match the station observations frequency and projected to the INCA domain. Although radar data is available every five minutes, every other observation was used to align with the temporal resolution of the station data and to reduce the weight on GPU memory.
Although radar data are available every 5~min, we used every second scan to match the 10~min station frequency and to reduce GPU memory usage.

MSG rapid-scan (5\,min) imagery was considered, though due to minimal skill improvement and doubled I/O cost, we retained the operational 15\,min channels for this initial implementation. To ensure consistency with the domain and 1km resolution of the radar and nowcasting data, the satellite data is cropped to the Swiss radar grid, interpolated accordingly, and interpolated linearly to a 10-minute temporal granularity.

\begin{figure}
    \centering
\includegraphics[width=.9\linewidth]{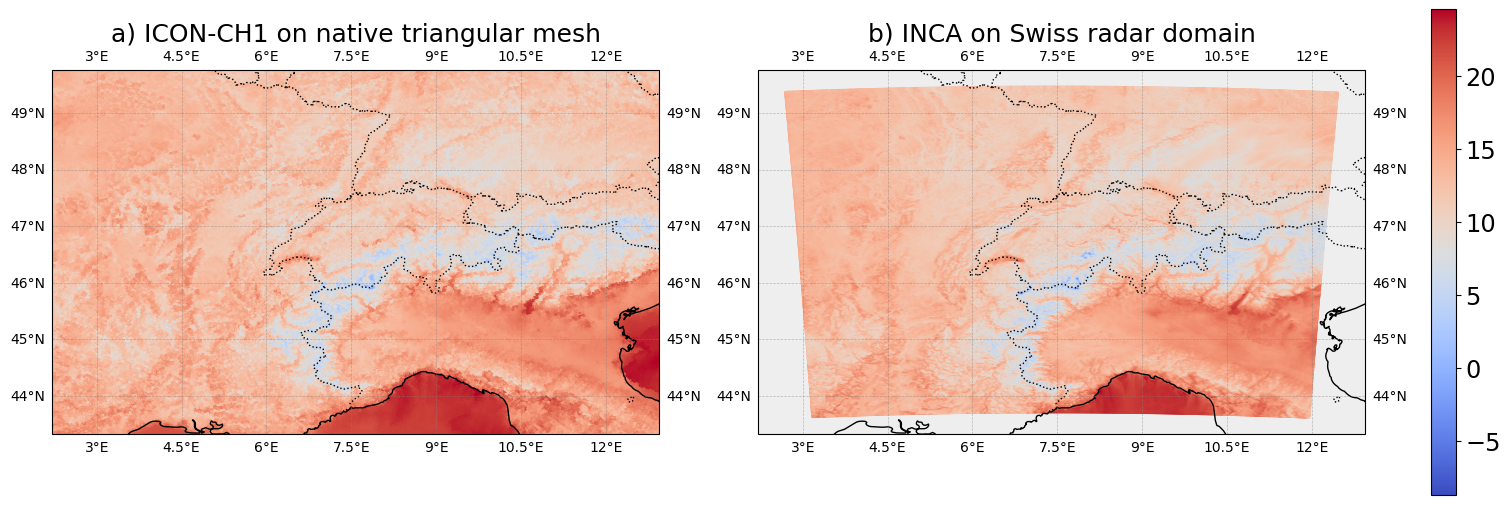}
\caption{2-meter temperature in $^{\circ}$C on the 27 Sept. 2023 at midnight for ICON--CH1 on its original triangular mesh (a) next to INCA nowcasting data for the same datetime (b). A triangular mesh is indexed with 1D edge coordinates, whereas a regular grid represents a set of areas of equal size formed by 2D coordinates.}
\label{icon}
\end{figure}

For the purpose of this study, the ICON--CH1 data on its original triangular mesh was reprojected on the Swiss radar 1km grid (EPSG:2056), Figure~\ref{icon}. The total accumulated precipitation is disaggregated to match the millimeter per hour unit of radar and station data. Temperature variables in Kelvin are converted into degrees Celsius to match INCA native units. ICON--CH1 data is then linearly interpolated in time to match the 10-minute nowcasting frequency.

\subsection{Predictors, Targets, Baselines, Ground truth}
We denote the five near-surface variables of interest as \texttt{U\_10M}, \texttt{V\_10M} (10\,m wind components; m\,s$^{-1}$), \texttt{TOT\_PREC} (rain rate; mm\,h$^{-1}$), \texttt{T\_2M}, and \texttt{TD\_2M} (2\,m air temperature and dewpoint; $^\circ$C). After spatial mapping to the Swiss radar grid and temporal coarsening/interpolation (Section~\ref{sec:temp_agg}), the working spatiotemporal resolution is uniformly 10~min and 1.11~km.

In order to make the predictor–target roles explicit for both AINCA and $\mathcal{L}$-AINCA (see Section~\ref{sec:experiments}), Table~\ref{tab:predictors} lists each data source, variables, spatio-temporal resolution, preprocessing steps, and its role in training and verification. ``Predictor" denotes a model input; ``target" a variable against which the training loss is computed; ``baseline", an operational forecast system used for comparison; and ``ground truth", a reference dataset treated as ground truth during verification. In Table~\ref{tab:predictors}, a source noted as both ``predictor'' and ``target'' means, in the context of forecasting, that the past data in $[t-1$h$,t]$ is used as input to predict $[t+10$min$,t+6$h$]$, with losses calculated against the corresponding targets evaluated in $[t+10$min$,t+6$h$]$.

We treat INCA analyses as the best available observation-informed estimate on the INCA grid. In practice, it means that scores are computed against INCA analyses during the quantitative verification. To assess generalisation, we also verify against a held-out subset of SwissMetNet stations (e.g., via pointwise metrics and meteograms).
\begin{table*}[t]
\centering
\caption{Predictors and targets.}
\label{tab:predictors}
\begin{tabular}{|p{1.4cm}|p{.9cm}|p{2.1cm}|p{1.5cm}|p{2.2cm}|p{1.6cm}|p{1.9cm}|}
\hline
\textbf{Source} & \textbf{Type} & \textbf{Variables}  & \textbf{Native resolution} & \textbf{Pre-processing} & \textbf{Role in AINCA} & \textbf{Role in $\mathcal{L}$-AINCA} \\
\hline
INCA & ana & \texttt{T\_2M}, \texttt{TD\_2M}, \texttt{U\_10M}, \texttt{V\_10M}, \texttt{TOT\_PREC} & 1.1~km, 10~min & --- & \textbf{target},  ground truth &  ground truth \\
\hline
INCA & fcst & \texttt{T\_2M}, \texttt{TD\_2M}, \texttt{U\_10M}, \texttt{V\_10M}, \texttt{TOT\_PREC} & 1.1~km, 10~min & --- & baseline & baseline \\
\hline
radar & obs. & \texttt{TOT\_PREC} & $\sim$1~km, 5~min & QC; clutter filter; linear interpolation to INCA grid & predictor & predictor, \textbf{target} \\
\hline
MSG SEVIRI & obs. & \texttt{IR\_108}, \texttt{IR\_016}, \texttt{VIS006}, \texttt{HRV}, \texttt{IR\_039} & $\sim$3~km, 15~min & resampled to 10min; linear interpolation to INCA grid & predictor & predictor \\
\hline
stations & obs. & \texttt{T\_2M, TD\_2M, U\_10M, V\_10M, TOT\_PREC} & point, 10~min & nearest neighbor interpolation to INCA grid & predictor, ground truth & predictor, ground truth, \textbf{target} \\
\hline
ICON--CH1 & fcst & \texttt{T\_2M, TD\_2M, U\_10M, V\_10M, TOT\_PREC} & $\sim$1.1~km, 1h & linear interpolation to INCA grid; unit conversion; disaggregation of precipitation; temporal slicing & predictor, baseline & predictor, baseline, \textbf{target} \\
\hline
\end{tabular}
\end{table*}

\subsection{Creation of the anemoi-dataset} 
Pre-processing steps, selected variables, date ranges, time frequency and spatial resolution are provided to \hyperlink{https://anemoi-datasets.readthedocs.io/en/latest/building/introduction.html}{anemoi-datasets} using a YAML configuration file and the command line interface (CLI) provided by the package. The data produced is sorted in a specific order so that slicing it results in 2D fields, and stored in zarr format. An additional wrapper providing a linkage of proprietary data is still necessary for internal sources, although anemoi-datasets provides support for many standard data formats (such as grib or netcdf files). The CLI also enables quick inspection of the built dataset, whereas opening a zarr file of this size (on the order of terabytes) using standard Python packages like \hyperlink{https://docs.xarray.dev/en/stable/user-guide/io.html}{xarray} might be very slow or even infeasible.

\subsection{Input tensor structure}
The network is fed with tensors covering the entire domain of interest ($710\times640$ grid cells). In contrast with the use of random square patches \citep[see, e.g.,][]{miralles2022downscaling}, the model might be able to learn the domain topography from weather variables which might alter out-of-sample performance or challenge the understanding of the physical aspects effectively learned by the model. However, training with square patches can result in difficulties reconstituting the full domain for verification and the use of arbitrary smoothing methods on the borders of the patches. We therefore favoured spatial seamlessness for this study, although it might be trickier to apply the model to other regions.

We use a 60-minute look-back window (six 10-minute steps) for observations (radar, stations, MSG) and only one present ($t$) and one future ($t+6$h) ICON--CH1 time step. The interpolator model ingests \begin{enumerate*}[i]
    \item the five satellite channels listed in Section~\ref{sec:data},
    \item the five near-surface variables at SwissMetNet stations, and
    \item radar-derived rain rate over the look-back window.
\end{enumerate*} 
In addition to observations, it also receives the five NWP surface variables at \(t\) and \(t{+}6\)h. An additional variable is introduced to give the GNN information about the target lead time position in the $[0, 6$h$]$ interval (e.g. 1/36 for lead time 10 minutes). Input tensors are of dimension three: the first dimension is the batch size (set to 1 in this study because model sharding is used, see Section~\ref{exp1}), the second is the flattened spatial coordinate ($454400$ points), and the last refers to the ``channels", i.e. individual meteorological variables $\times$ individual time steps per variable ($5\times2+11\times6+1=77$ input channels in total). All predictors are normalised using their mean and standard deviation.

\subsection{Data splits and loss calibration}
The training period spans from August 2023 to July 2024, which represents about 52600 samples ($\approx$ 85~\% of the total available data). Validation and test data are built from subsequent months, respectively, August and September 2024 (containing about 4000 initialisation times each). Aggregate metrics are based only on the independent test period (labeled ``Test aggregate” in the figure caption), while case studies (labeled ``Case study” in the figure caption) are explicitly flagged and may include training data.

The decay factor and weights in the total loss (Equation~\ref{loss_exp2}) were set to $\alpha=0.15, w_1=0.4$, $w_2=0.4$ and $w_3=0.2$ after a qualitative calibration of the hyperparameters, where different weight combinations were manually evaluated by inspecting the model outputs over the validation set (similar to a model validation phase) to identify a configuration that yielded the best results.

\section{Computational Footprint}
\label{exp1}
 The GNN has about $110$ million parameters and is trained in $\sim$120h on 16 nodes with 4$\times$A100~(40~GB) GPUs, an AMD~EPYC 64-core CPU, and 512~GB RAM. Parallel data loading is implemented directly through Anemoi. At the time of this study, Anemoi supported model sharding, which is parallel training of model partitions, only with a batch size of 1. Because sharding accelerated convergence more than increasing the batch size, we used a batch size of 1. Convergence is carefully monitored; the validation loss appears to plateau after about $35$~epochs (e.g. the training takes $\sim3.5$h per epoch).
 
A comprehensive hyperparameter calibration was not feasible given the limited GPU budget. Instead, during training of $\mathcal{L}$ -AINCA, we adaptively updated the loss weight per source of \eqref{loss_exp2}, while varying the length of the Fourier transform signal had a negligible impact on performance.

Inference is performed to generate the full 12-hour sequences, with a temporal resolution of 10 minutes. Each 12-hour sequence takes approximately 25 seconds to compute, which corresponds to roughly 0.35 seconds per 10-minute timestep on a single GPU. For comparison, INCA requires 5 (temperature variables)–10 (surface wind) min per surface-variable run, yielding an approximate 12–24$\times$ speedup.

In terms of storage needs, each checkpoint is about $40$GB while the anemoi-dataset used for training is $1.4$TB.
The total size of forecasts over the test set stored in zarr format is around $65-70$GB per model. The baseline forecasts from ICON--CH1 and INCA over the test set and the reference analyses are about $50$GB each, which amounts to a total verification footprint of roughly $300$GB. 

\section{Verification}
\label{sec:verif}
\subsection{Scores definition}
The Fractions Skill Score \citep{roberts2008scale} for threshold $\tau$ and window size $s$ can be defined as 
\begin{equation}
\mathrm{FSS}(\tau,s)
= 1 - 
\frac{\displaystyle \sum_{\omega \in \Omega_s} \big(\Po(\omega) - \Pf(\omega)\big)^2}
     {\displaystyle \sum_{\omega \in \Omega_s} \big(\Po(\omega)^2 + \Pf(\omega)^2\big)},
\label{eq:fss-def-rl}
\end{equation}
with $\Po$ the proportion of observed values and $\Pf$ the proportion of forecasted values above the threshold $\tau$. In this study, we use various values for threshold $\tau$ and a window size of $s=10$km mainly to capture intense precipitation events.

We use the Root Mean Squared Error (RMSE) as defined in \citet{hyndman2006another} and the following formula for the Pearson correlation:
\begin{equation*}
    \rho
  = \frac{\displaystyle\sum_{i=1}^{n} \left(f_i - \overline{f}\right)\,\left(o_i - \overline{o}\right)}
          {\sqrt{\displaystyle\sum_{i=1}^{n} \left(f_i - \overline{f}\right)^2}\;
           \sqrt{\displaystyle\sum_{i=1}^{n} \left(o_i - \overline{o}\right)^2}} \,
\end{equation*}
where $f$ are model forecasts, $o$ ground truth values and $\overline{f}, \overline{o}$ their respective average over the test set.

\subsection{Quantitative: Scores against analysis}
Scores are calculated for the test set, defined in Section~\ref{sec:data_dl}. We select 240 initialization times (one every 3 hours) within the test set, run a 6-hour sequence, compute the score against INCA analysis, and compare to baseline forecast systems. The model can be run for 12-hour sequences, as it encodes the relative lead time with respect to the forecast horizon and adjusts predictions accordingly. Nevertheless, since training was conducted on a 6-hour horizon, quantitative scores are reported for 6-hour sequences to ensure methodological consistency and fair comparison. For qualitative event-based verification, however, 12-hour sequences are used, as relevant weather events typically extend beyond 6 hours.
\begin{figure}[!ht]
    \centering
\includegraphics[width=\linewidth]{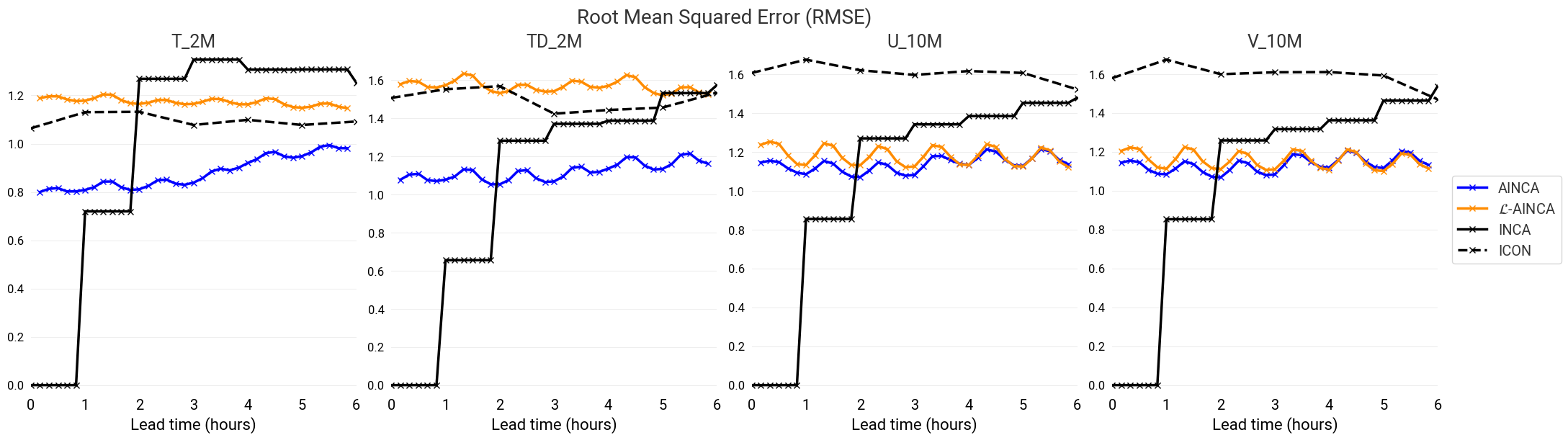}
\includegraphics[width=\linewidth]{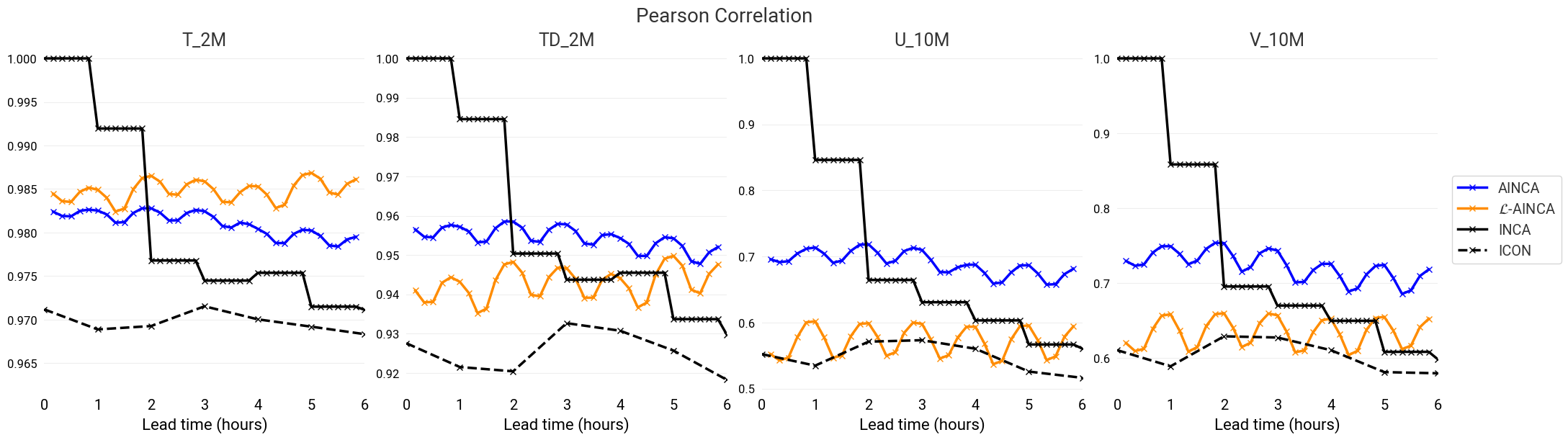}
\includegraphics[width=\linewidth]{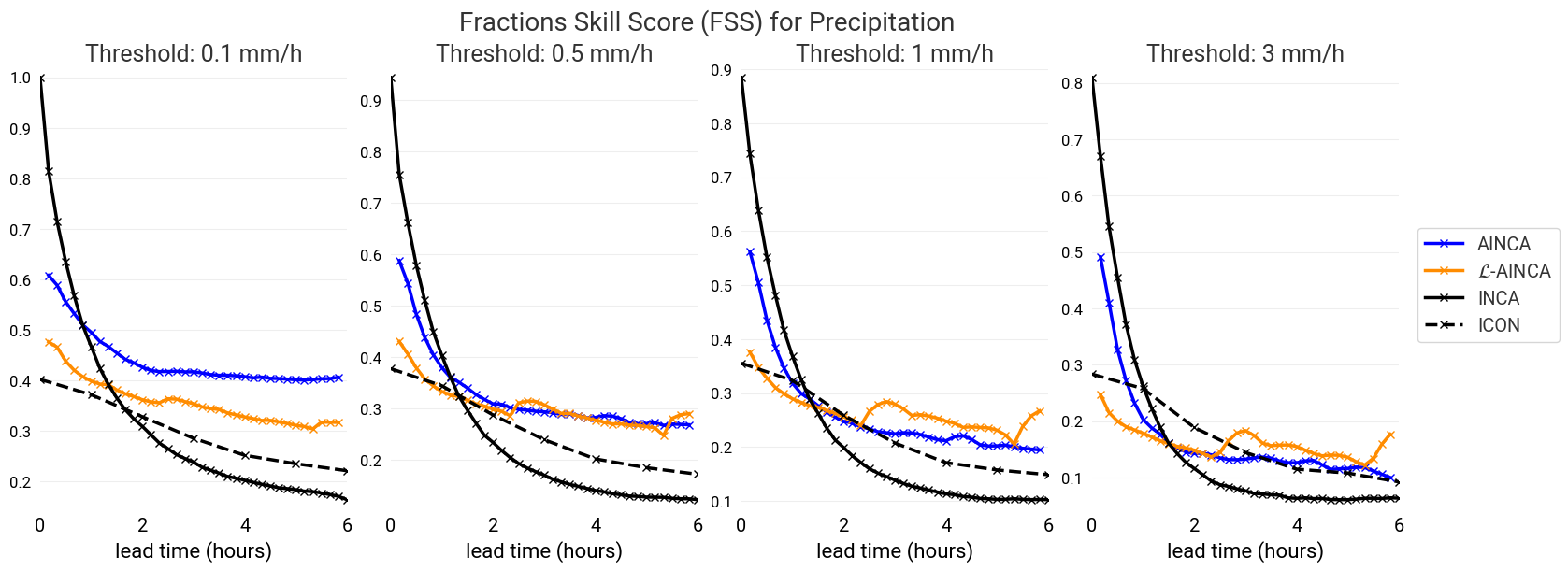}
\caption{Average root mean squared error, Pearson correlation, Fraction skill score for rain rate for thresholds 0.1, 0.5, 1 and 3 mm.h$^{-1}$ over 10km spatial windows. Scores are computed versus INCA analysis; ICON--CH1 and INCA forecasts are used as baselines for comparison. Test set aggregate.}
\label{fig:scores}
\end{figure}
Some scores vary a lot with pointwise discrepancies (e.g., the root mean squared error), whereas some others are designed to focus on spatial patterns (e.g., the fractions skill score). Figure~\ref{fig:scores} shows both types of scores for models resulting from AINCA and $\mathcal{L}$-AINCA (see Section~\ref{sec:experiments}), evaluated against INCA analysis data, which is considered the ground truth and, as such, is referred to as ``target" in some of the plots. 
Scores for the ICON--CH1 (black continuous line) and INCA (black dotted line) forecasts are also shown for comparison. 
The rain rate was not directly available from the ICON--CH1 forecast output, but was derived by computing the difference in total accumulated precipitation between two consecutive lead times as explained in Section~\ref{sec:data_dl}. 

Figure~\ref{fig:scores} provides several insights. First, by design, INCA forecasts match the analysis at $t_0$ and gradually blend into ICON--CH1 by about $+6$ hours. Nevertheless, we observe a small discrepancy in skill, which may arise from using ICON--CH1 runs different than those used operationally in INCA. Indeed, INCA runs using the most recent available NWP data, which can lag up to one hour behind the latest update. Second, both data-driven models, AINCA and $\mathcal{L}$-AINCA, outperform ICON--CH1 across most variables, metrics, and lead times. At very short lead times, however, data-driven forecasts cannot reproduce or substitute the analysis and are therefore outperformed by INCA forecasting system. 
For longer lead times ($\geq 2$h), AINCA consistently improves upon the INCA forecasting system for all variables and scores against INCA analysis, while $\mathcal{L}$-AINCA model yields lower RMSE than the INCA forecasts for wind components. Part of this trend may reflect the effect of INCA forecasts blending into ICON–CH1. Finally, AINCA (dark blue) was trained to emulate INCA analyses, which implies that RMSE at $t_0$ should be close to zero. Interestingly, the model sacrifices exact matching at time zero in favor of temporal consistency.

For precipitation rate, conventional pointwise metrics such as RMSE and Pearson correlation are less meaningful due to the \textit{double penalty} effect, where small spatial or temporal shifts in precipitation are counted as both misses and false alarms. Instead, we report the fractions skill score (computed over 10~km patches at different thresholds in mm.h$^{-1}$), which explicitly accounts for spatial tolerance. Figure~\ref{fig:scores} show that AINCA and $\mathcal{L}$-AINCA consistently outperform INCA rain rate forecasts for lead times $\geq$40 minutes.
\begin{figure}[!ht]
    \centering
    \includegraphics[width=\linewidth]{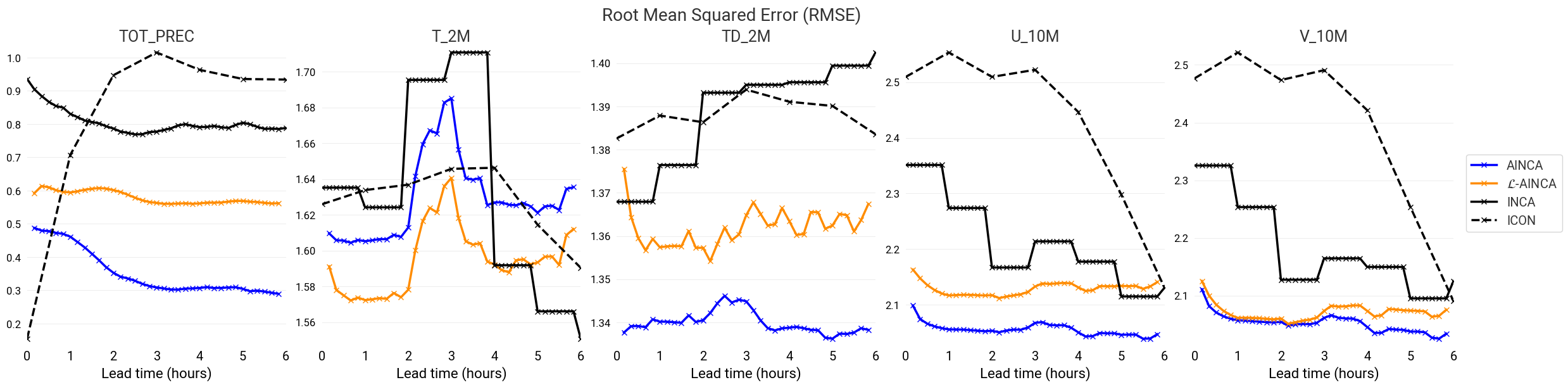}
    \caption{Spatio-temporal average root mean squared error versus 2200 partner stations not used in training of AINCA or $\mathcal{L}$-AINCA over the test period. ICON--CH1 and INCA forecasts are used as baselines for comparison. Test set aggregate.}
\label{fig:scores_obs}
\end{figure}

When scores are computed against the 2200 partner stations held out from training (Section~\ref{subsec:station}), the short-lead advantage of the INCA forecast system is not observed (Fig.~\ref{fig:scores_obs}). Both AINCA and $\mathcal{L}$-AINCA outperform the baselines at all lead times for the wind components, rain rate, and dewpoint temperature. For 2-m temperature, they also surpass INCA for lead times smaller than 4 hours.
For \texttt{T\_2M}, a mid lead-time RMSE bump (near 3h in Figure~\ref{fig:scores_obs}) is visible. It could be linked with strong diurnal cycles and sunrise/evening transitions. Figure~\ref{fig:scores_obs} also shows particular patterns for wind variables \texttt{U\_10M} and \texttt{V\_10M} in RMSE versus station observations with baseline models scores decreasing with lead times. It might be because very short–lead winds contain small–scale turbulence and local channeling. 

\begin{figure}[!ht]
    \centering
    \includegraphics[width=0.62\linewidth]{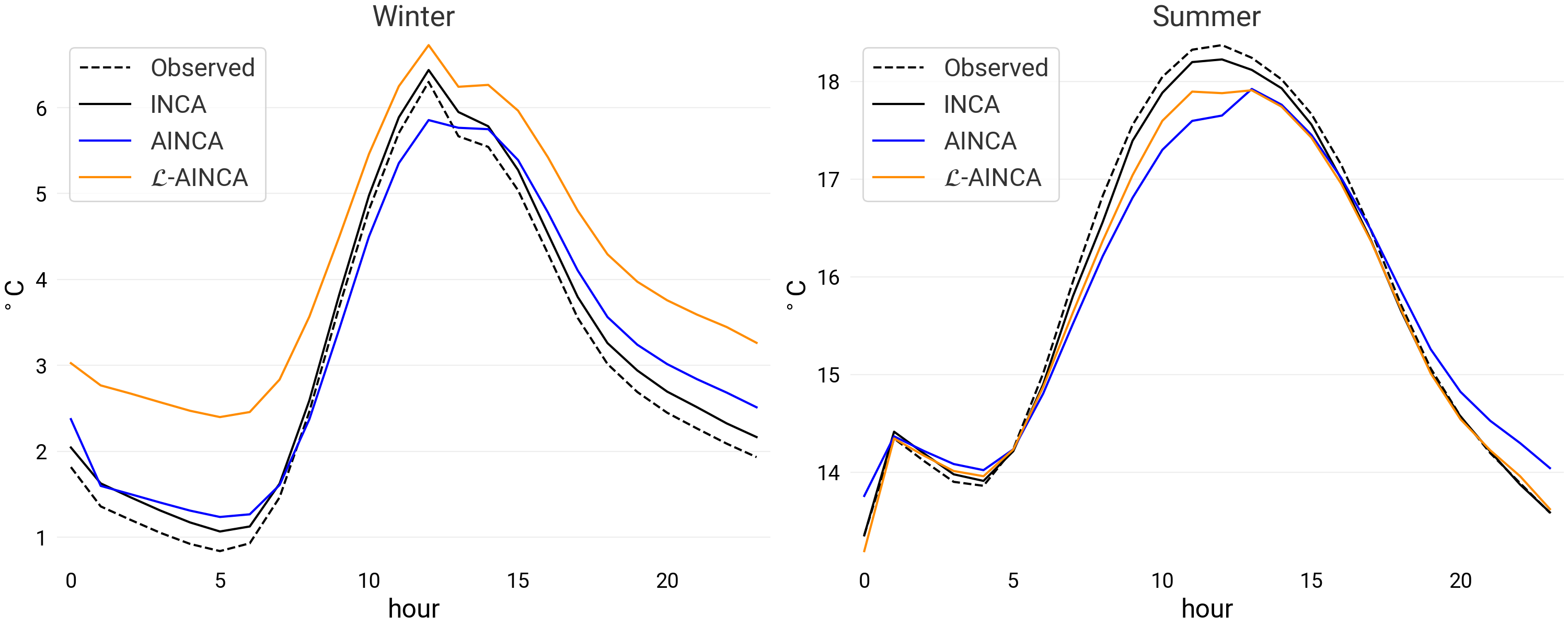}
    \includegraphics[width=0.37\linewidth]{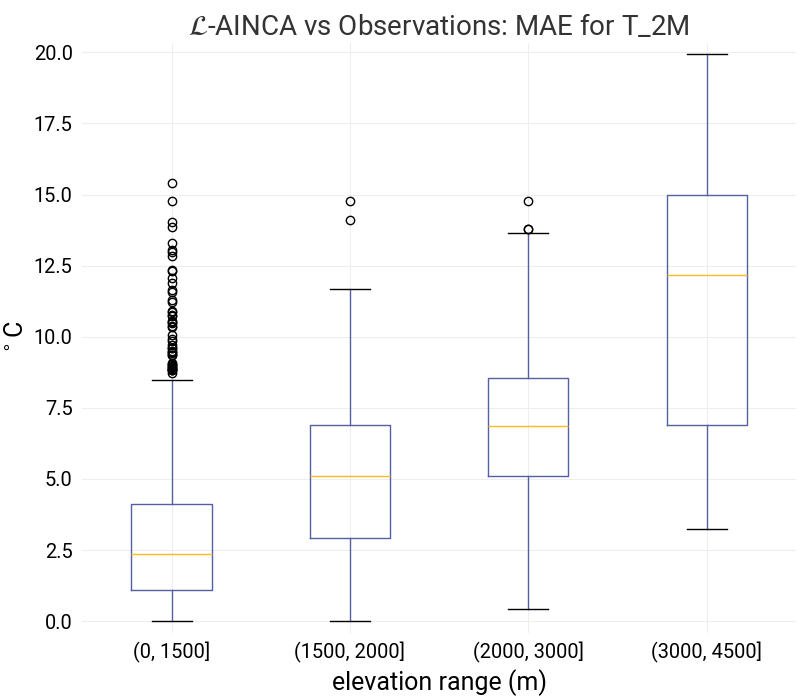}
    \caption{Spatio-temporal mean daily pattern for predicted air temperature \texttt{T\_2M} in winter and summer compared to analysis (black) and observations (black dotted). Boxplot of the mean absolute error in winter between $\mathcal{L}$-AINCA predictions and observations, grouped by elevation bins. Note that the bins contain unequal numbers of stations: 1268, 172, 126, and 14 respectively. No significant variation in the \texttt{T\_2M} MAE distribution is observed for stations located between 0 and 1500 meters elevation. Case study.} 
    \label{fig:mdp}
\end{figure}
Interestingly, the GNN performs worst against INCA analysis for temperature variables (Figure~\ref{fig:scores}), despite their relatively high predictability from topography and diurnal cycle. As shown in Figure~\ref{fig:mdp}, $\mathcal{L}$-AINCA systematically overestimates winter temperatures, with the largest departures from the mean daily cycle at high-altitude stations. This likely reflects the difficulty of capturing Alpine local processes. Indeed, cold nights are related to temperature inversions (stratus or fog), cold air pooling, snow cover, and related feedbacks that are poorly captured by ICON--CH1. In contrast, warm nights are typically associated with large-scale cloudy conditions and advected air masses, which are comparatively easier to forecast.

\subsection{Significance testing}
\label{sec:sig}
We report the average RMSE, mean bias and Pearson correlation computed on the INCA 1.1~km grid at each forecast reference time from the test set sliced every 3hours. For precipitation, we additionally report the Fractions Skill Score (FSS) using thresholds $\{0.05, 0.1, 1, 3\}$~mm.h$^{-1}$ and 10~km neighborhood windows.

In this study, we chose not to display confidence intervals on the raw score plots. The uncertainty of individual scores is typically dominated by weather variability and therefore very large, which makes it uninformative for assessing relative forecast quality. Predictability fluctuates with the weather and the scores vary accordingly; therefore, to compare two forecasts in situations of varying predictability, the relevant quantity is the uncertainty of the difference in scores between forecasts rather than the uncertainty associated with each score in isolation. Some visualizations of scores uncertainty over the horizon of interest are provided in the appendix (\ref{fig:boxplots}).

We only report uncertainty and significance in the context of score differences, using INCA and ICON--CH1 as the baseline models. Pairwise differences between systems (AINCA, $\mathcal{L}$-AINCA, INCA, ICON) are assessed with paired Diebold-Mariano tests for each variable, model and benchmark.

Diebold-Mariano corrected two-sided test statistic \citep{Hering01112011} compares a pair of forecasts by computing the mean and estimating the standard deviation of the difference in scores between a forecast and a baseline model. The skill score is then defined as the improvement ratio, e.g. the mean difference divided by the average score of the baseline model, or
\begin{equation}
\mathrm{SS} = \begin{cases}
    \dfrac{\mathrm{S}_{\mathrm{ref}}-\mathrm{S}_{\mathrm{fcst}}}{\mathrm{S}_{\mathrm{ref}}} & \text{if the score is of type ``lower is better" (ex: RMSE)} \\
   \dfrac{\mathrm{S}_{\mathrm{fcst}}-\mathrm{S}_{\mathrm{ref}}}{\mathrm{S}_{\mathrm{ref}}} & \text{if the score is of type ``higher is better" (ex: FSS)}, 
\end{cases}
    \label{eq:skill_score}
\end{equation}
where $\mathrm{S}_{\mathrm{ref}}$ is the baseline score and $\mathrm{S}_{\mathrm{fcst}}$ the model score. The two-sided test then determines whether the mean difference between the scores of a baseline model and a DL model is statistically significant with confidence 95\%.
\begin{figure}[!ht]
    \centering
\includegraphics[width=.8\linewidth]{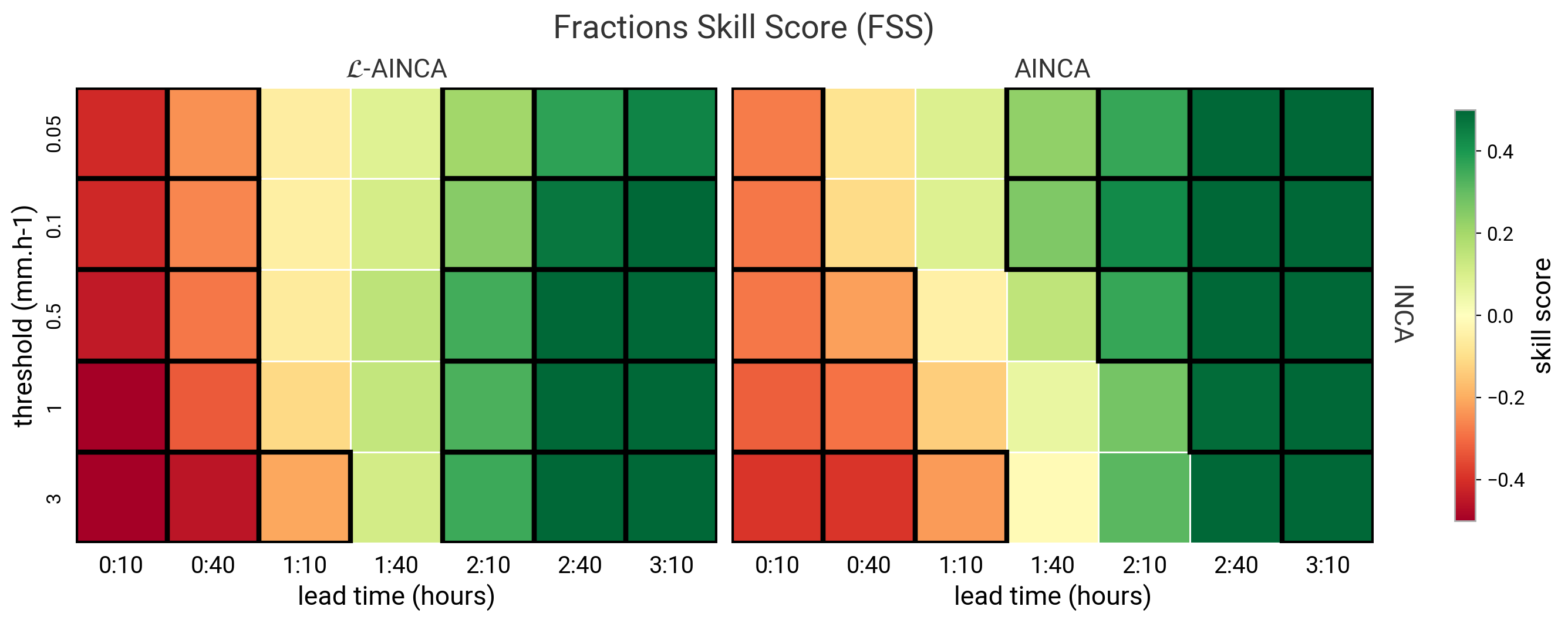}
\includegraphics[width=.49\linewidth]{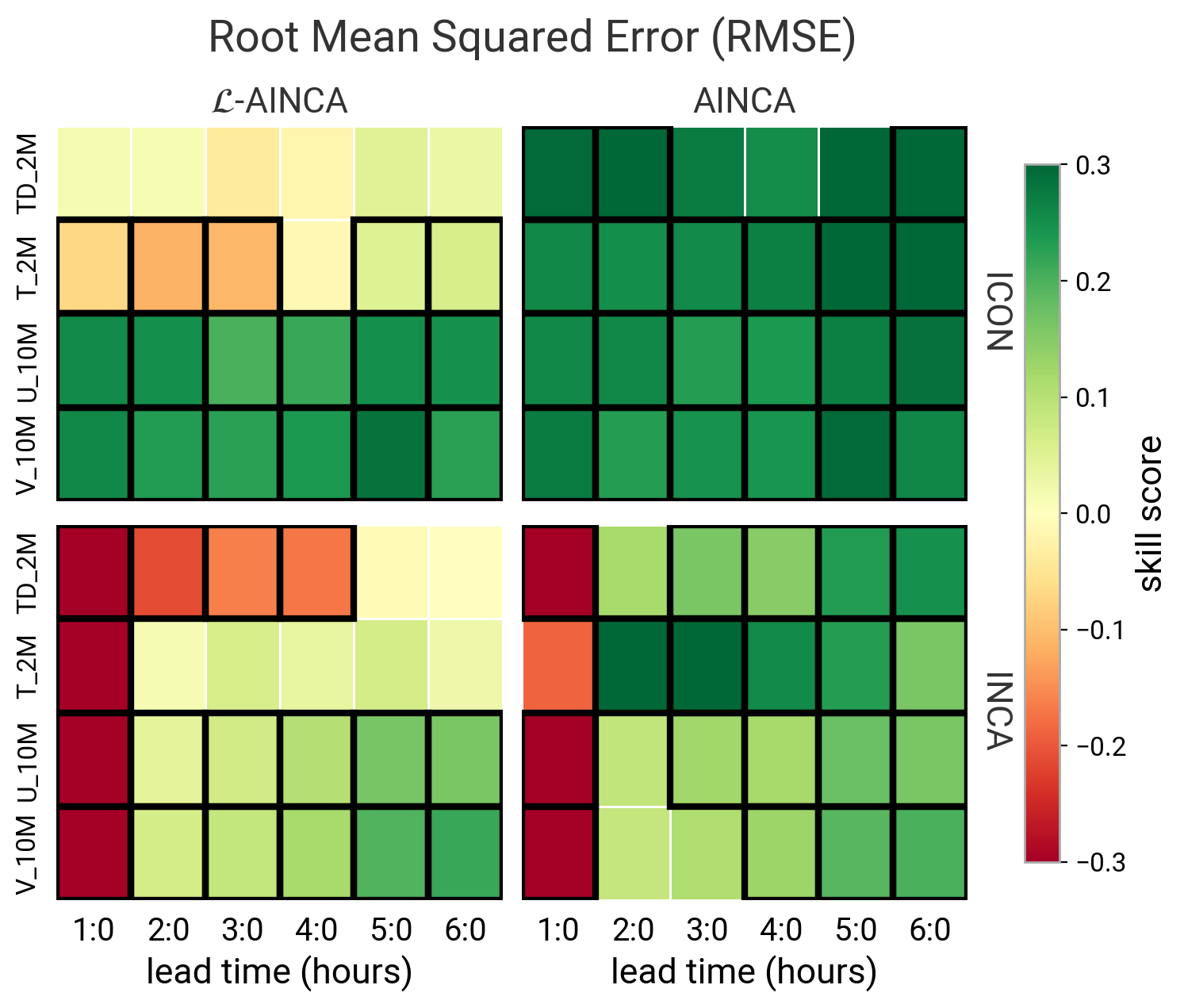}
\includegraphics[width=.49\linewidth]{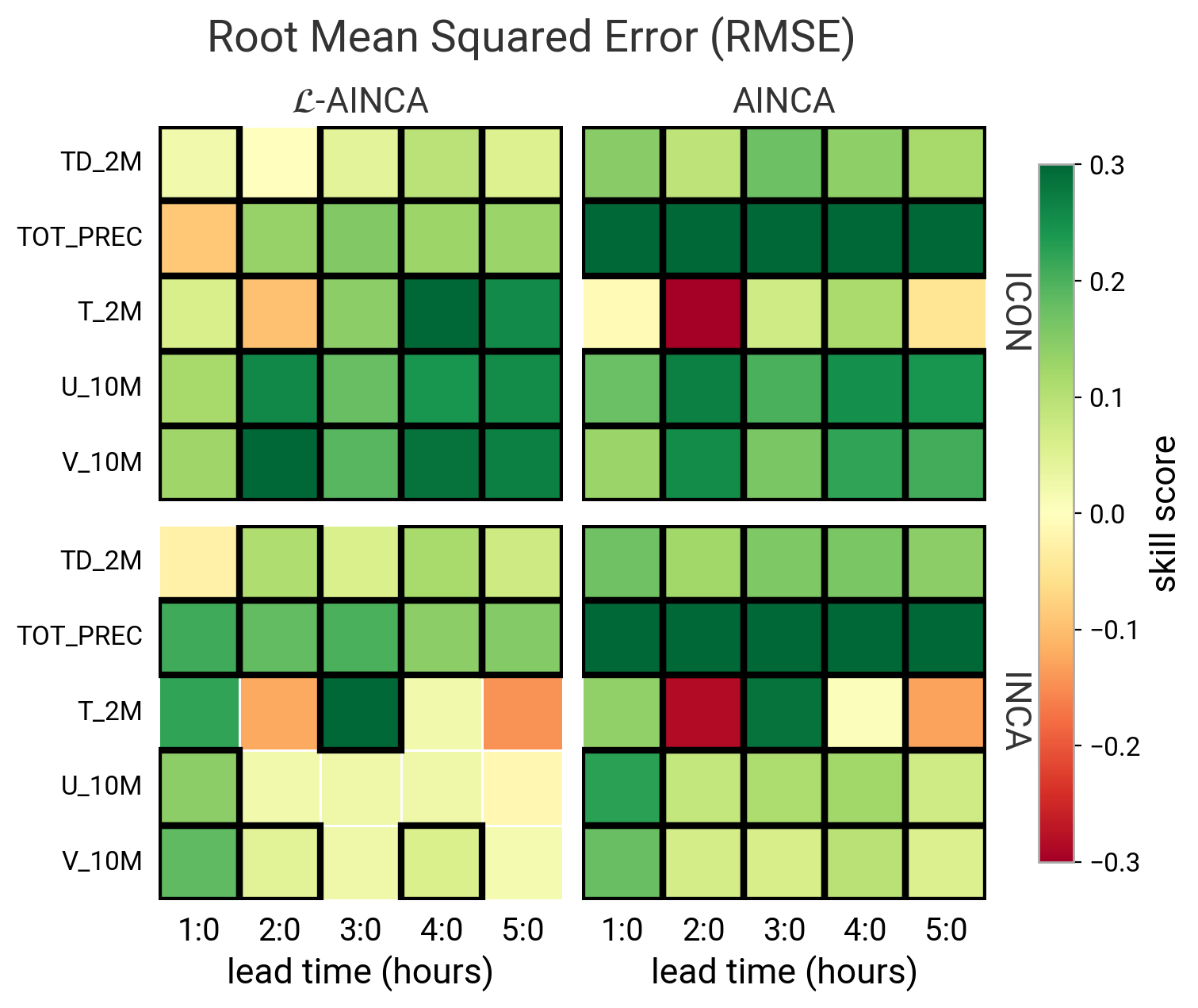}
    \caption{Skill score (Equation~\ref{eq:skill_score}) for fractions skill score against INCA analysis (top) and RMSE (bottom) against INCA analysis (left) and partner stations observations (right). ICON--CH1 and INCA forecasts are used as baselines for comparison (rows) with DL models (columns). Red and green boxes show respectively negative and positive skill score for each DL model compared to the baseline. Thick borders mean the difference is statistically significant according to the Diebold-Mariano statistic. Test set aggregate.}
\label{fig:dm_test}
\end{figure}

Figure~\ref{fig:dm_test} summarizes the pairwise comparisons: the \emph{upper} panel shows the fractions skill score of the rain rate against INCA analysis, while the \emph{lower} panels report the RMSE differences against the INCA analysis (left) and against the partner stations (right).

Against the INCA analysis (lower left), both AINCA and $\mathcal{L}$-AINCA beat ICON--CH1 for \texttt{U\_10M} and \texttt{V\_10M} at all lead times and surpass INCA beyond 1h. For \texttt{T\_2M} and \texttt{TD\_2M}, AINCA is better than ICON--CH1 at all lead times and exceeds INCA after 1h; $\mathcal{L}$-AINCA shows no consistent gain for these two variables. The early-lead disadvantage of the DL models relative to INCA is expected: INCA forecasts are warm-started from the INCA analysis at $t_0$, whereas the DL models are not conditioned on that analysis.

Verification against stations (lower right) corroborates this conjecture. AINCA and $\mathcal{L}$-AINCA significantly outperform ICON--CH1 in most variables and lead times, with only isolated lead time / variable pairs significantly favoring ICON--CH1. AINCA also exceeds INCA for the vast majority of cases (all but \(2/25\)). $\mathcal{L}$-AINCA clearly outperforms INCA for rain rate, dewpoint temperature, and winds; for \texttt{U\_10M}, gains beyond 1h are positive but not always statistically significant.

\subsection{Qualitative: Event-based}
\begin{table}[!ht]
    \centering
    \begin{tabular}{|p{2.2cm}|p{2cm}|p{2cm}|p{3.3cm}|p{3cm}|}
    \hline
       \textbf{Variable} & \textbf{Start date} & \textbf{End date} & \textbf{Description} & \textbf{Stations} \\ \hline
         \texttt{U\_10M V\_10M} & 2023-02-25 & 2023-02-26 & Bise situation & KLO, WAE, BER, QUI, SMA, GRA, PAY \\ \hline
         \texttt{U\_10M V\_10M T\_2M} & 2023-03-12 & 2023-03-13 & Stormy front over Swiss Plateau & CRM, NEU, PAY, MUB, CHA, CHM \\ \hline
         \texttt{U\_10M V\_10M} & 2023-03-30 & 2023-03-31 & Storm Mathis & KOP, BER, BAN, GRE, LAG, WYN \\ \hline
        U\_10M V\_10M & 2023-06-20 & 2023-06-21 & Dynamic trough leading an active cold front & GVE, CGI, DOL, PRE, BIE \\ \hline
        \texttt{TOT\_PREC} & 2023-08-26 & 2023-08-29 & Hail event in locarno, afterwards heavy precipitation & BIA, LOM \\ \hline
        \texttt{U\_10M V\_10M} & 2024-02-24 & 2024-02-25 & South foehn event & ALT, GES, ENG, GLA, CHU \\ \hline
        \texttt{U\_10M V\_10M} & 2024-03-01 & 2024-03-02 & South foehn event & ALT, GES, ENG, GLA, CHU \\ \hline
        \texttt{U\_10M V\_10M} & 2024-03-08 & 2024-03-09 & South foehn storm & GOR, ZER, SIM, MTE, EVO \\ \hline
        \texttt{U\_10M V\_10M} & 2024-03-24 & 2024-03-26 & South foehn event & ALT, GES, ENG, GLA, CHU \\ \hline
        \texttt{U\_10M V\_10M} & 2024-06-21 & 2024-06-22 & North foehn combined with thunderstorm outflow & PIO, BIA, CEV, COM, MAG, OTL, LUG \\ \hline
        \texttt{U\_10M V\_10M TOT\_PREC} & 2024-06-28 & 2024-06-29 & Intense thunderstorms cause flooding in Valle Maggia & ALT, ULR, SIO, MTR, BIA, LUG \\ \hline
        \texttt{U\_10M V\_10M} & 2024-09-10 & 2024-09-14 & North foehn event & PIO, COM, BIA, CEV, GRO, MAG, LUG, SBO \\ \hline
    \end{tabular}
    \caption{Meteorological events occurring during the data availablity period. Stations are specified by their SwissMetNet short name. Case study.}
\label{tab:events}
\end{table}
Scores alone are insufficient to assess the plausibility of the generated forecasts, so we also visualize key variables for selected events, listed in Table~\ref{tab:events}. Unlike in quantitative verification, visualizations can also be qualitatively evaluated for events in the training set because no sign of overfitting was detected in the quantitative verification, and many interesting events occurred in the training period.
\begin{figure}[!ht]
    \centering
\includegraphics[width=\linewidth]{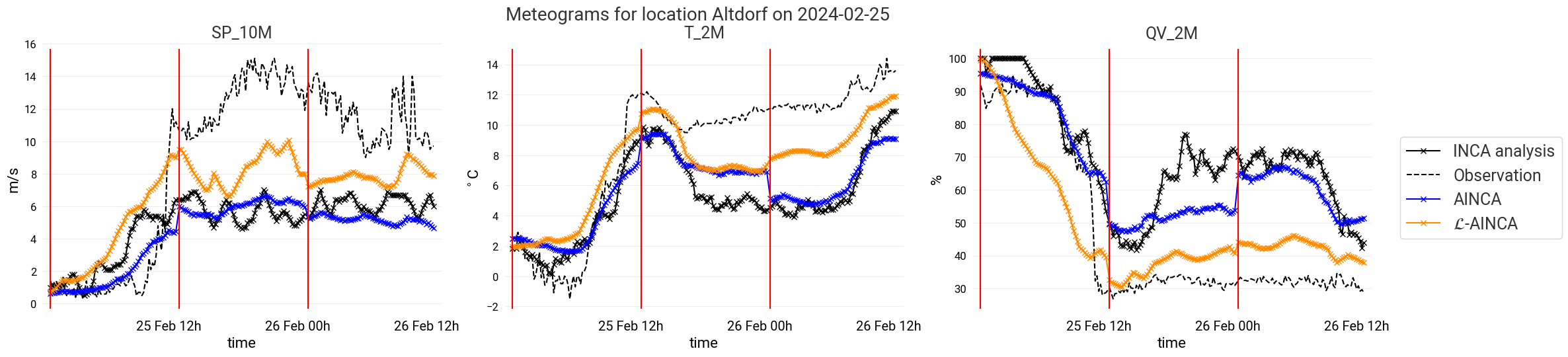}
\includegraphics[width=\linewidth]{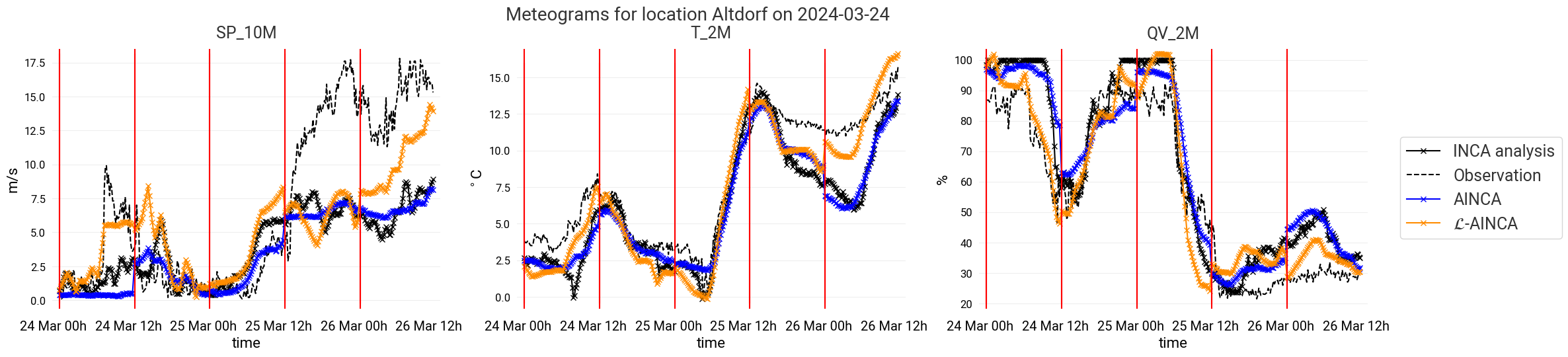}
\caption{Meteograms for the 25 February (top) and 24 March (bottom) 2024 South Foehn events at Altdorf (ALT). The first column represents the 10-meter wind speed in m.s$^{-1}$, the second 2-meter temperature, and the last shows relative humidity in \%. Events are further described in Table~\ref{tab:events}. Red vertical lines show successive initialisation times for the AINCA and $\mathcal{L}-$AINCA forecasts.}
\label{fig:meteograms}
\end{figure}
Meteograms presented in Figure~\ref{fig:meteograms} illustrate how AINCA produces smoother patterns than $\mathcal{L}$-AINCA. During the South Foehn events that occurred on the 24/25 February and 24 March 2024, both AINCA and $\mathcal{L}$-AINCA models successfully captured key features to be expected at a site in the lee of the mountain range: The foehn onset is characterized by often sharp increases in temperature and wind speed, along with a drop in relative humidity. Altdorf, known for its exposure to Foehn events, serves as a valuable reference point. While the two models are not perfectly aligned, the temporal evolution of the individual variables during Foehn onsets (shortly before 00:00 UTC on 25 February and in the morning of the 24 March 2024) is fairly consistent in each of the models and agrees with the physical expectation. Some differences can also be noted between the measured observations and the INCA analysis, which looks smoother. As shown in Figure~\ref{fig:meteograms}, the $\mathcal{L}$-AINCA model tends to produce temporal variations that are more similar to those of the observations than the AINCA model or the analysis data, especially with regards to wind speed.

In addition to event-specific meteograms, time series of maps throughout the day played a key role during validation. Some models (here meant as checkpoint) achieving better quantitative scores exhibited unrealistic spatial patterns or artifacts, in particular for the rain rate, and were discarded. Instead, hyperparameters were adjusted to prioritize models that produced more plausible spatiotemporal behavior. Some illustrative precipitation patterns are shown in Figures~\ref{fig:examples_precip}, and further examples of predicted temperature, dewpoint, and wind patterns over 12 hours are shown in the appendix (Figures~\ref{fig:examples_temp} and~\ref{fig:examples_wind}).

\section{Conclusion}
\label{sec:conclusion}
In this study, we build an end-to-end pipeline that integrates heterogeneous data sources in an operational setting. The use of the maintained codebase of Anemoi and ECMWF was essential: without it, the engineering effort to develop and test data-ingestion and sampling pipelines and deep-learning models would be prohibitively costly for smaller weather services, even given that open-source codebases for deep learning models exist. Moreover, using Anemoi for data preprocessing, training, and inference ensures that the model architecture and pipeline are transparent and reproducible. This approach supports adaptation to other regions or forecasting systems and remains open for collaborative development.

This study shows promising improvements employing graph neural networks for high-resolution nowcasting in topographically complex regions such as Switzerland compared to traditional nowcasting methods. The main benefits come from the computational efficiency achieved at inference time, the capacity of a single model to jointly predict all surface variables, including precipitation rate at 10-minute intervals, and the overall improvements observed across standard verification metrics for lead times larger than 40min against INCA analysis and for all lead times against station observations.

Using observational data, numerical weather prediction inputs, and a flexible loss function tailored to the nowcasting context, the model can produce forecasts with competitive accuracy and spatial coherence without the need for existing nowcasting analyses. Furthermore, the inference speed is well within practical limits for deep learning-based forecasting, and significantly faster than traditional numerical models, making it suitable for both research and near-real-time applications.

Quantitative evaluations show that the GNN models outperform ICON--CH1 forecasts against INCA analysis over the test set accross all lead times and variables. Compared to the operational nowcasting system INCA, scores computed against INCA analysis show improvement using the DL models for lead times larger than 2 hours. Scores computed against station observations show that DL models both mostly outperform ICON--CH1 and INCA forecasts over the 6 hour horizon, and do not show any particular under-performance of the DL models for short lead times. Qualitative assessments in the form of visual pattern inspections show that the model output is often indistinguishable from traditional analyses, supporting its operational viability.

Despite these encouraging results, challenges remain. The temperature variables proved unexpectedly difficult to predict accurately. Furthermore, the limited availability of extreme events in the training data can restrict the generalizability of the model in high-impact situations. Furthermore, while the model implicitly handles several types of uncertainty, explicit uncertainty quantification remains an open research direction.

GNN-based nowcasting offers a path forward for localised, high-resolution short-term weather forecasting. Future work should explore the integration of ensemble methods for uncertainty estimation, expanding training datasets to better capture extreme events, and evaluating transferability to other regions.

 \acknowledgments
Computational resources were provided by the Swiss National Supercomputing Centre (CSCS) and the Federal Office of Meteorology and Climatology (MeteoSwiss), which also provided data. The work benefited from fruitful discussions with the SEPP team at MeteoSwiss and helpful feedback from Pr. Anthony Davison (EPFL). 

 \datastatement 
 This study was designed with reproducibility and transferability in mind. The GNN model and training pipeline build on the open source Python package Anemoi and the forks used for this study are available on GitHub (\href{https://github.com/OpheliaMiralles/anemoi-training}{github.com/OpheliaMiralles/anemoi--training}, \href{https://github.com/OpheliaMiralles/anemoi-models}{github.com/OpheliaMiralles/anemoi--models}). SwissMetNet station data can be downloaded freely from the MeteoSwiss website (\href{https://www.meteoswiss.admin.ch/services-and-publications/service/open-data.html}{www.meteoswiss.admin.ch/services--and--publications/service/open--data.html}), satellite data is available via the EUMETSAT API (\href{https://data.eumetsat.int}{data.eumetsat.int}), composite radar data can be obtained from the MeteoFrance API (\href{https://portail-api.meteofrance.fr/web/fr/api/RadarOpera}{portail-api.meteofrance.fr/web/fr/api/RadarOpera}) and topographic data can be downloaded freely from the SRTM 90m DEM Digital Elevation Database (\href{http://srtm.csi.cgiar.org}{srtm.csi.cgiar.org}). EUMETSAT, OPERA and DEM data can alternatively be downloaded and stored in zarr format using the open source centralized data platform \hyperlink{https://github.com/MeteoSwiss/weathermart/blob/main/README.md}{github.com/MeteoSwiss/weathermart} developed by MeteoSwiss. Forecast and analysis archive data from the ICON--CH1 and INCA model and from the MeteoSwiss radar are not open-source but can be obtained from MeteoSwiss on demand. MeteoSwiss data is increasingly available as Open Government Data (see \href{https://opendatadocs.meteoswiss.ch/}{opendatadocs.meteoswiss.ch}).

\contributionstatement
O.M. conducted the research and wrote the manuscript. B.R. implemented software components critical to nowcasting experiments in Anemoi. J.B. provided critical feedback on the manuscript. D.N. contributed to the research direction. C.S. offered suggestions for extending the work. All co-authors reviewed and approved the final version.

\clearpage
\appendix
\section{Illustrative Examples: Forecasts vs INCA analysis}
We present animations of surface variables, decomposed to illustrate the forecast evolution over time in comparison with the ground truth and the baseline. 

\begin{figure}[!ht]
    \centering
\includegraphics[width=\linewidth]{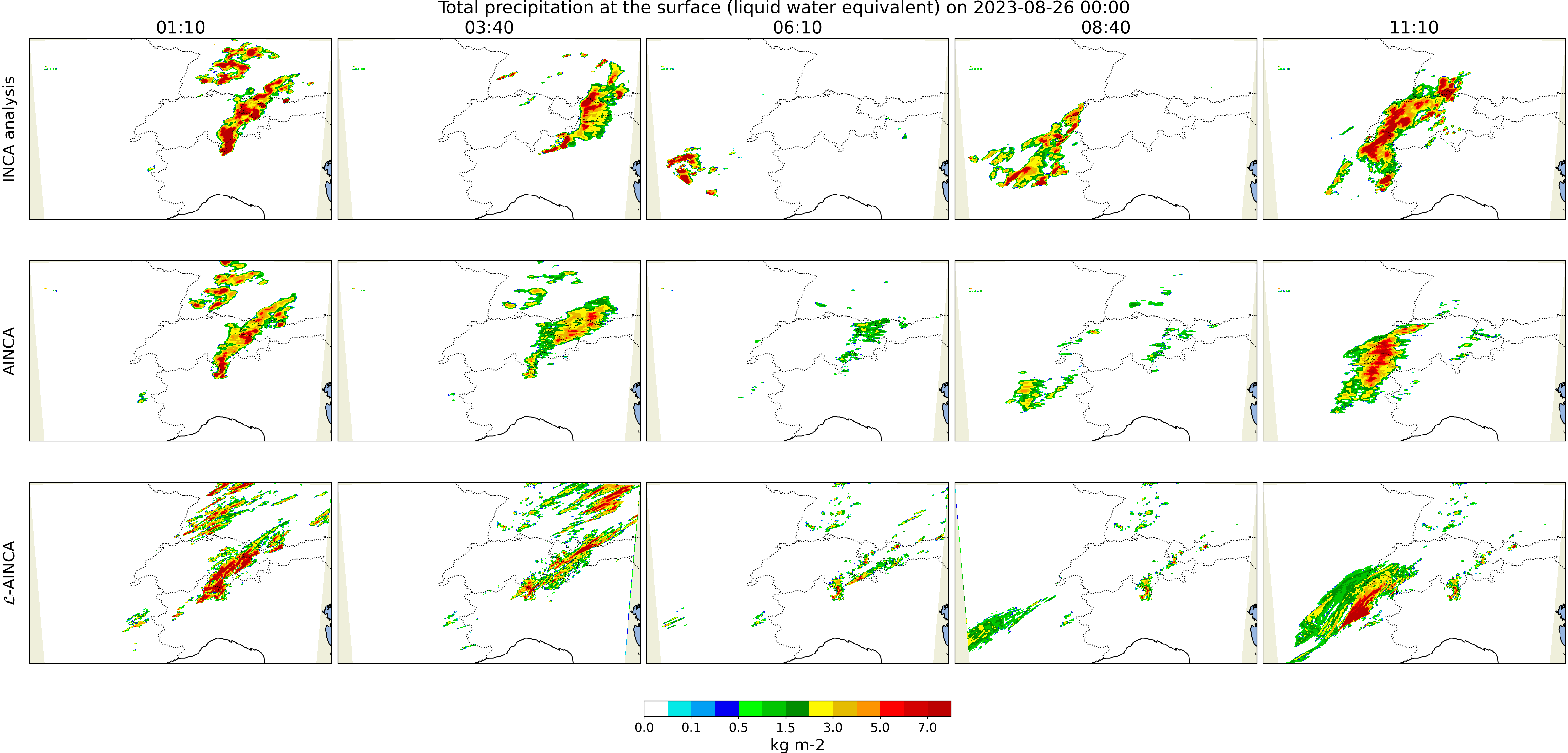}
\caption{Example image sequence for instantaneous precipitation (rain rate). Case study.}
\label{fig:examples_precip}
\end{figure}

Precipitation patterns, such as those illustrated in Figure~\ref{fig:examples_precip}, are notoriously difficult for deep learning models to reproduce. These models often behave like linear interpolators, producing overly smooth or lagged outputs that lack physical realism. In contrast, our results demonstrate that instantaneous precipitation is advected across the domain in a manner closely resembling the analysis.

\begin{figure}[!ht]
    \centering
\includegraphics[width=\linewidth]{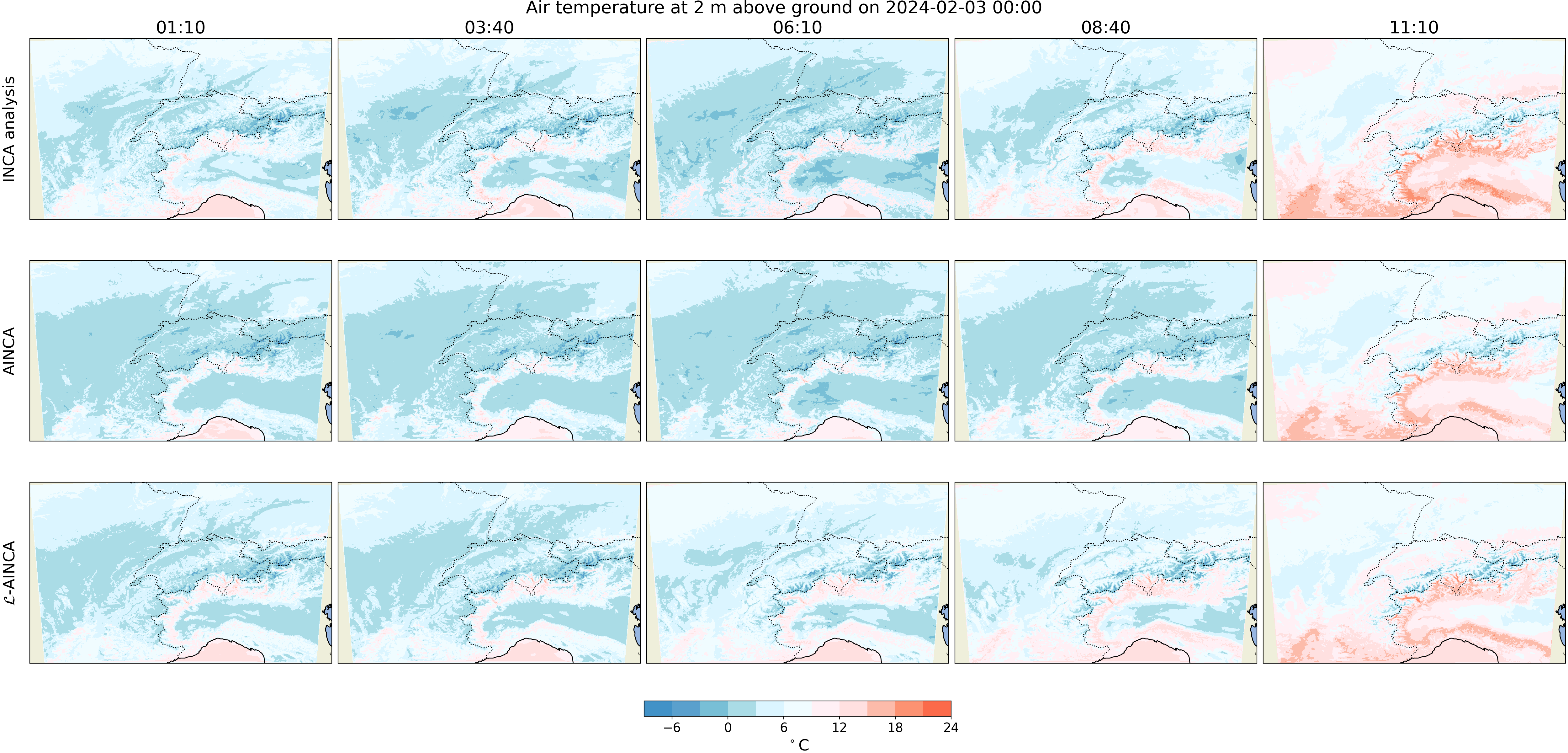}
\includegraphics[width=\linewidth]{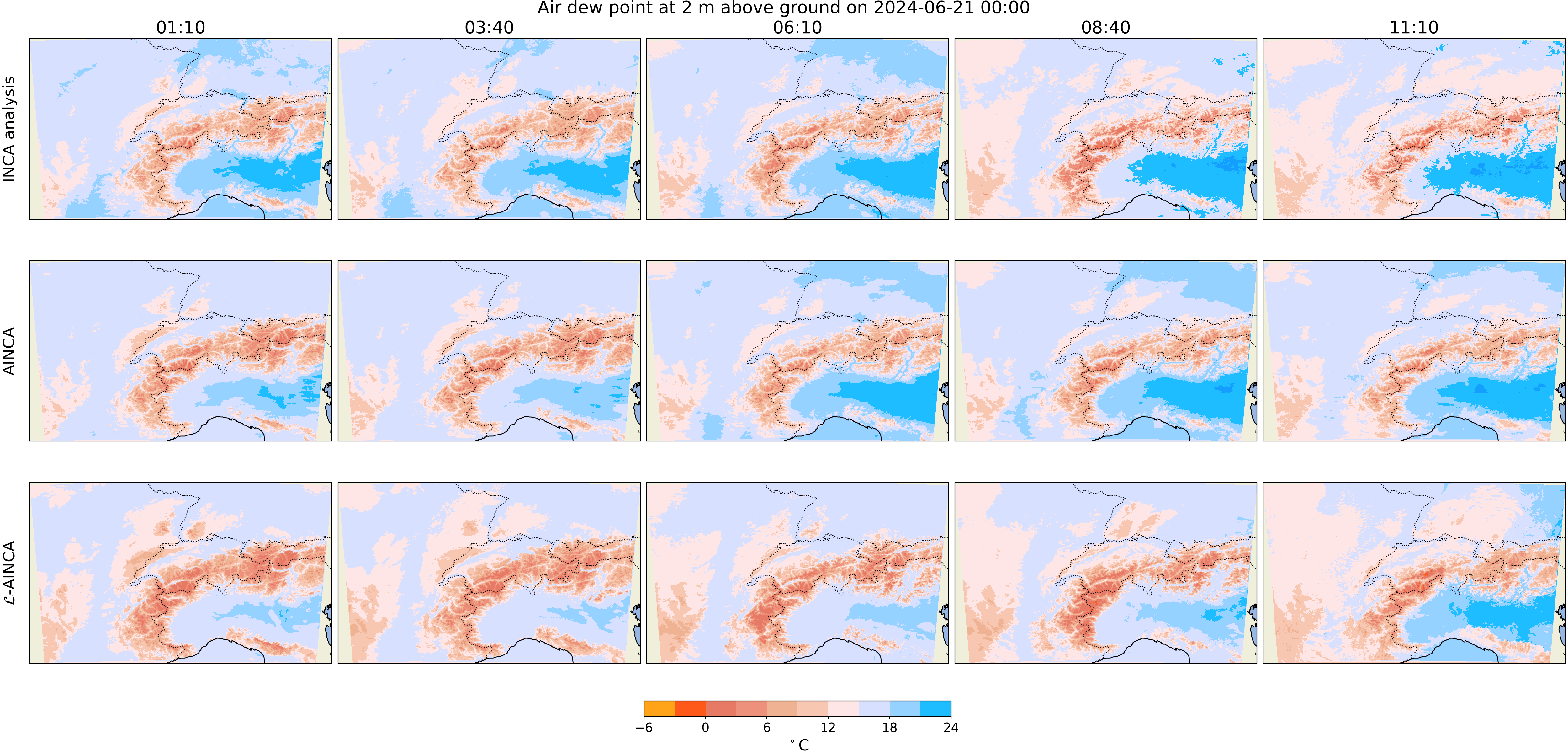}
\caption{Example image sequences for 2-meter temperature and 2-meter dewpoint temperature. Case study.}
\label{fig:examples_temp}
\end{figure}

\begin{figure}[!ht]
    \centering
\includegraphics[width=\linewidth]{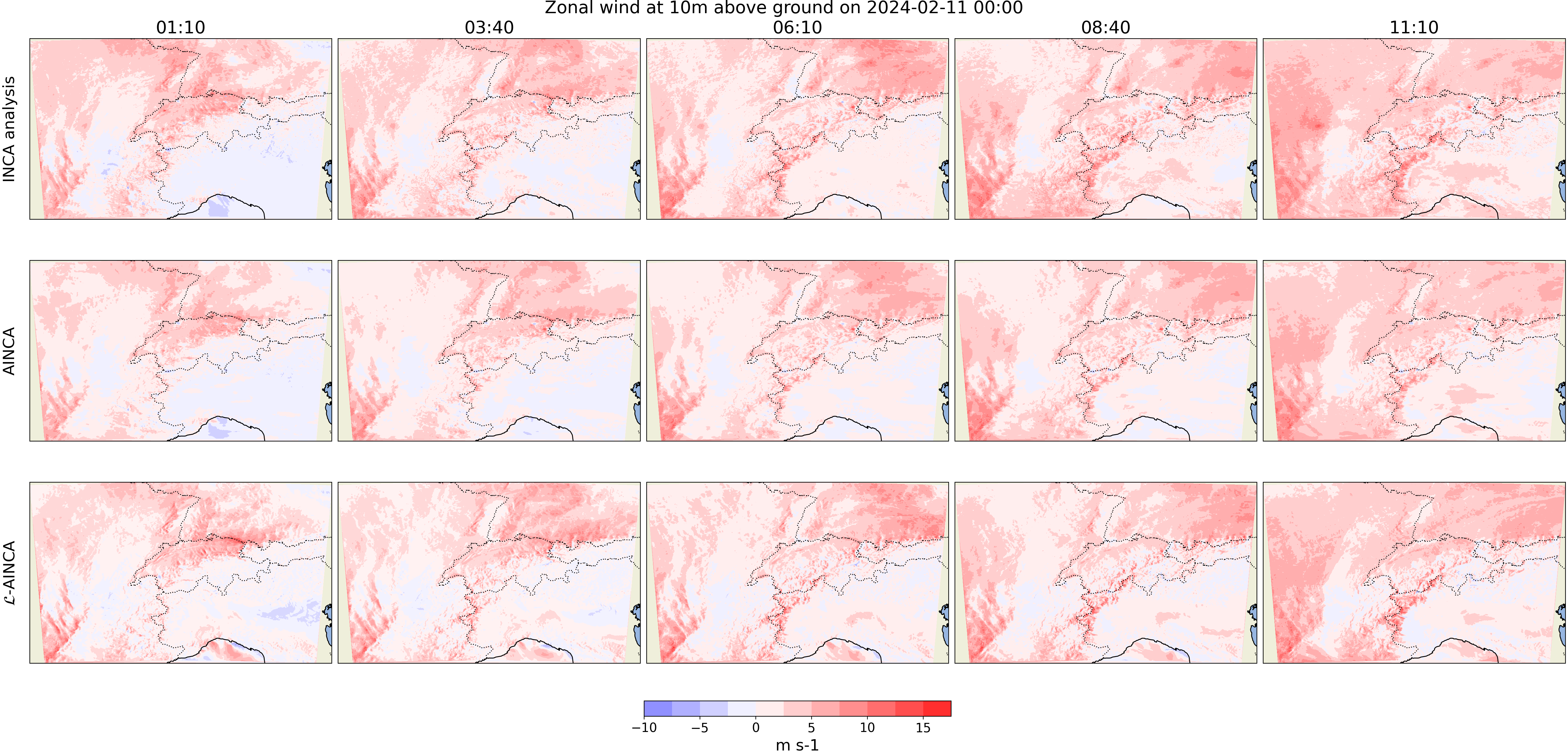}
\includegraphics[width=\linewidth]{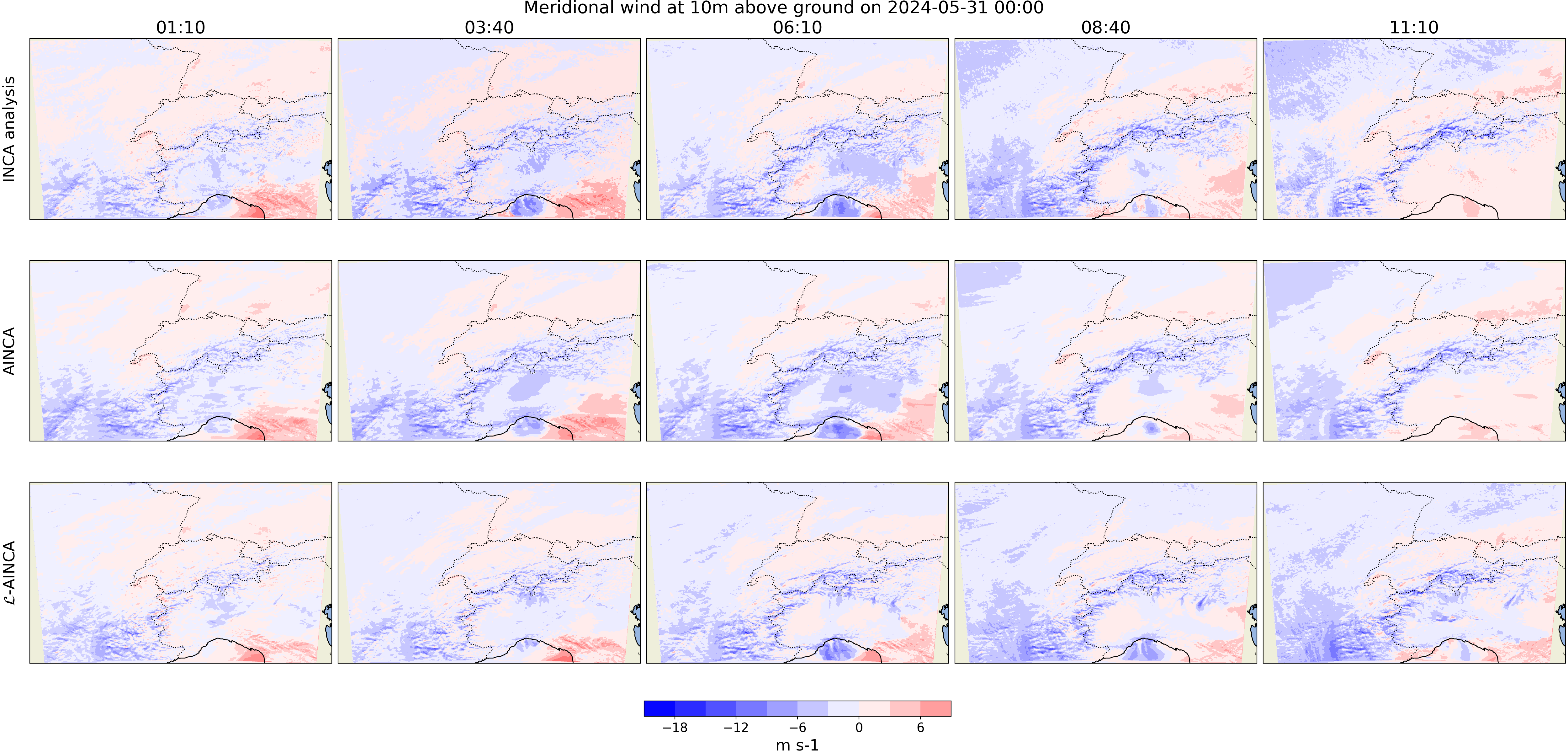}
\caption{Example image sequences for 10-meter northern and eastern wind components. Case study.}
\label{fig:examples_wind}
\end{figure}

The following examples for 2-meter temperature/dewpoint temperature (Figure~\ref{fig:examples_temp}) and 10-meter wind components   (Figure~\ref{fig:examples_wind}) show that 12-hour time series of predicted values closely resemble the INCA analysis data.

\clearpage
\section{Scores distribution: Boxplots}
Although we do not include uncertainty intervals on the main score plots, we still consider it useful to give a complementary view of the score distribution across the test set. Here we provide boxplots of the scores for each model and baseline for 1,$\cdots$,6hours lead times for wind and temperature variables, and up to 3hours lead time for rain rate. 

\begin{figure}[!ht]
    \centering
    \includegraphics[width=\linewidth]{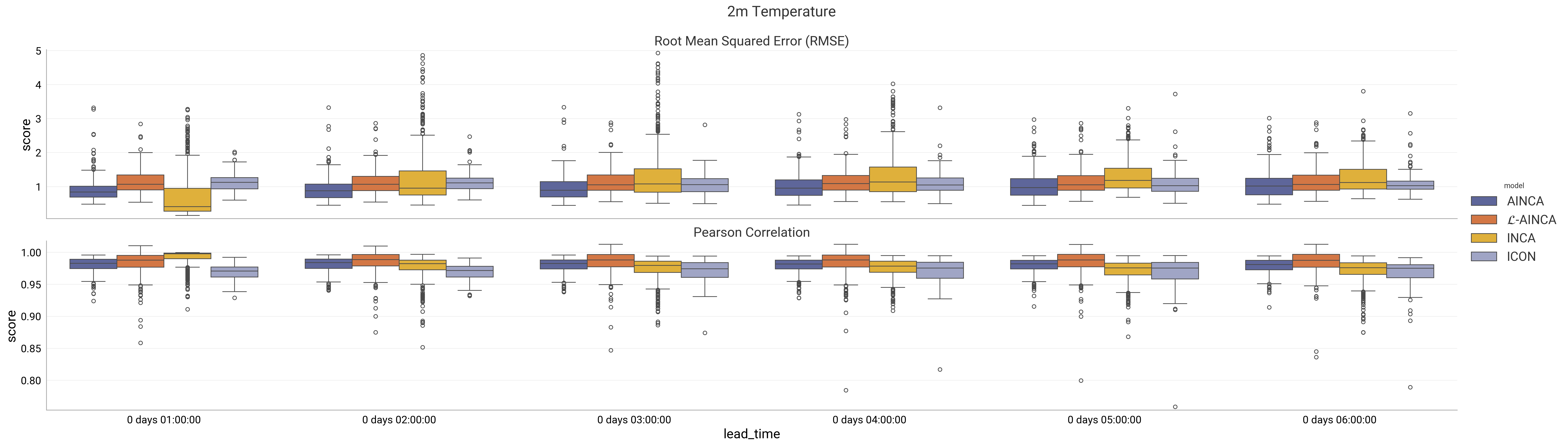}
    \includegraphics[width=\linewidth]{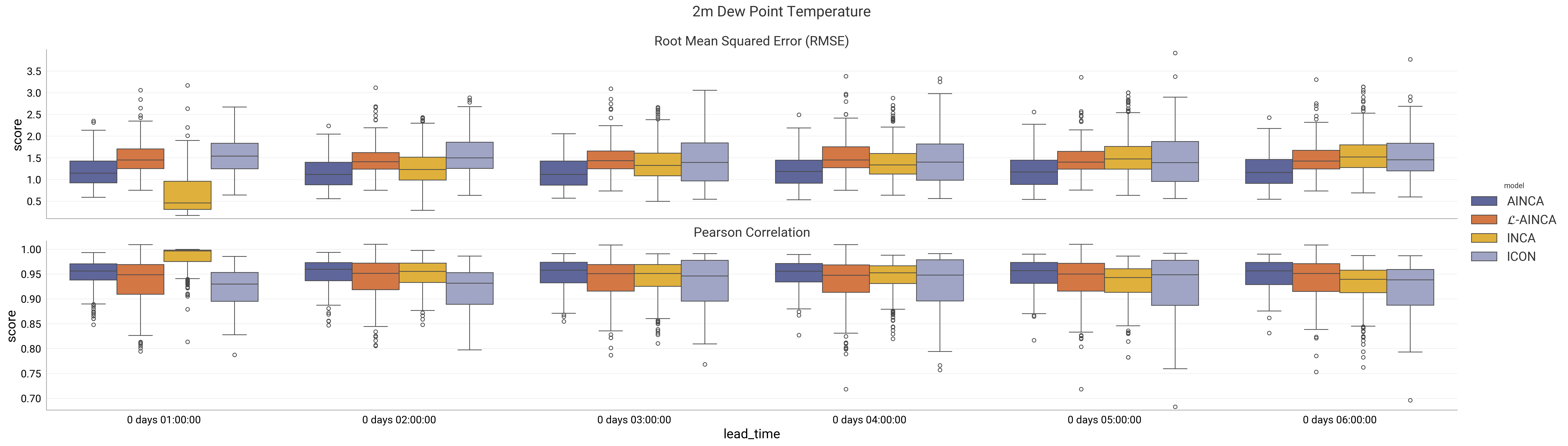}
\includegraphics[width=\linewidth]{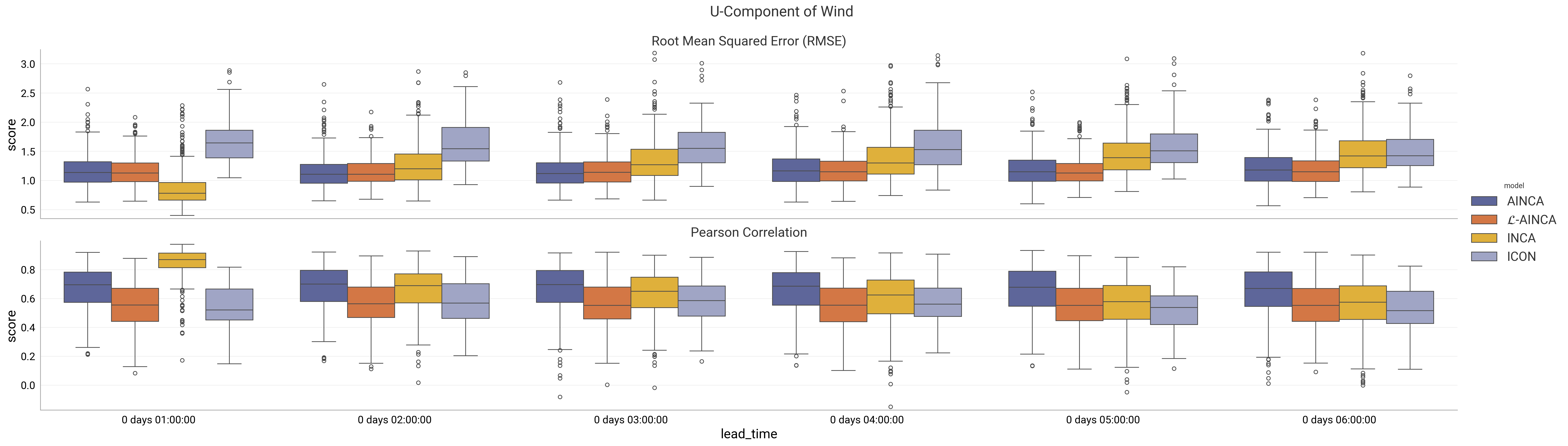}
\includegraphics[width=\linewidth]{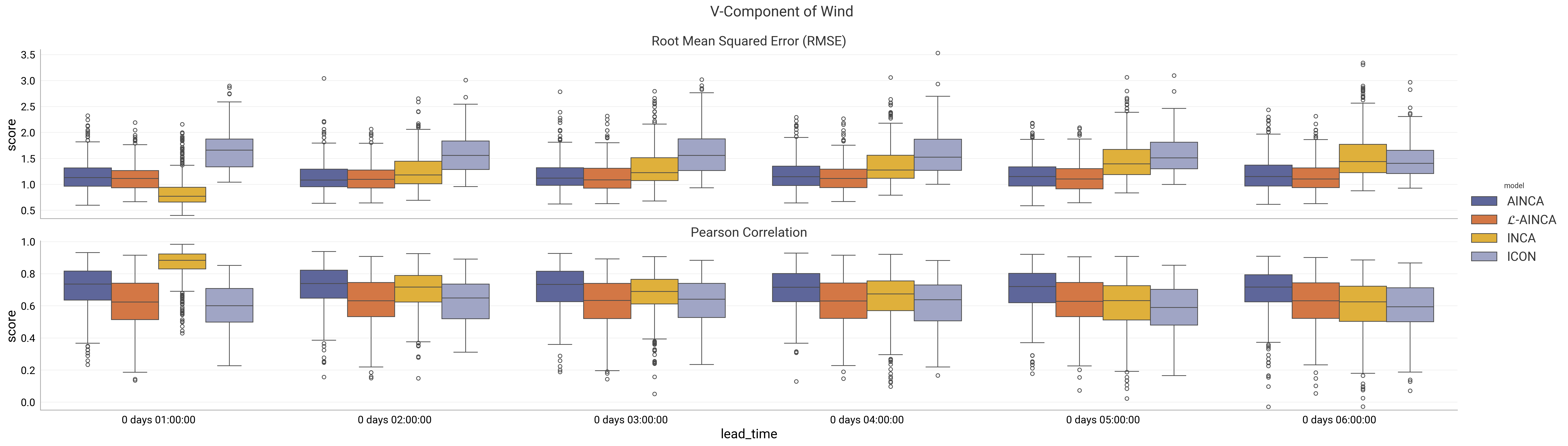}
\caption{Boxplot time series of the score distribution over the test set for temperature and wind variables. Test aggregate.}
\label{fig:boxplots}
\end{figure}

Figure~\ref{fig:boxplots} shows that AINCA and $\mathcal{L}$-AINCA exhibit considerably less variability across the test set compared to the INCA nowcasting system, likely reflecting the spatial smoothing characteristic of deep learning models. From lead times of 2 hours onward, AINCA consistently exceeds the baseline models, with lower upper bounds in RMSE and higher upper bounds in Pearson correlation. The $\mathcal{L}$-AINCA model shows similarly low variability but, as expected, performs slightly worse.
\begin{figure}[!ht]
    \centering
    \includegraphics[width=\linewidth]{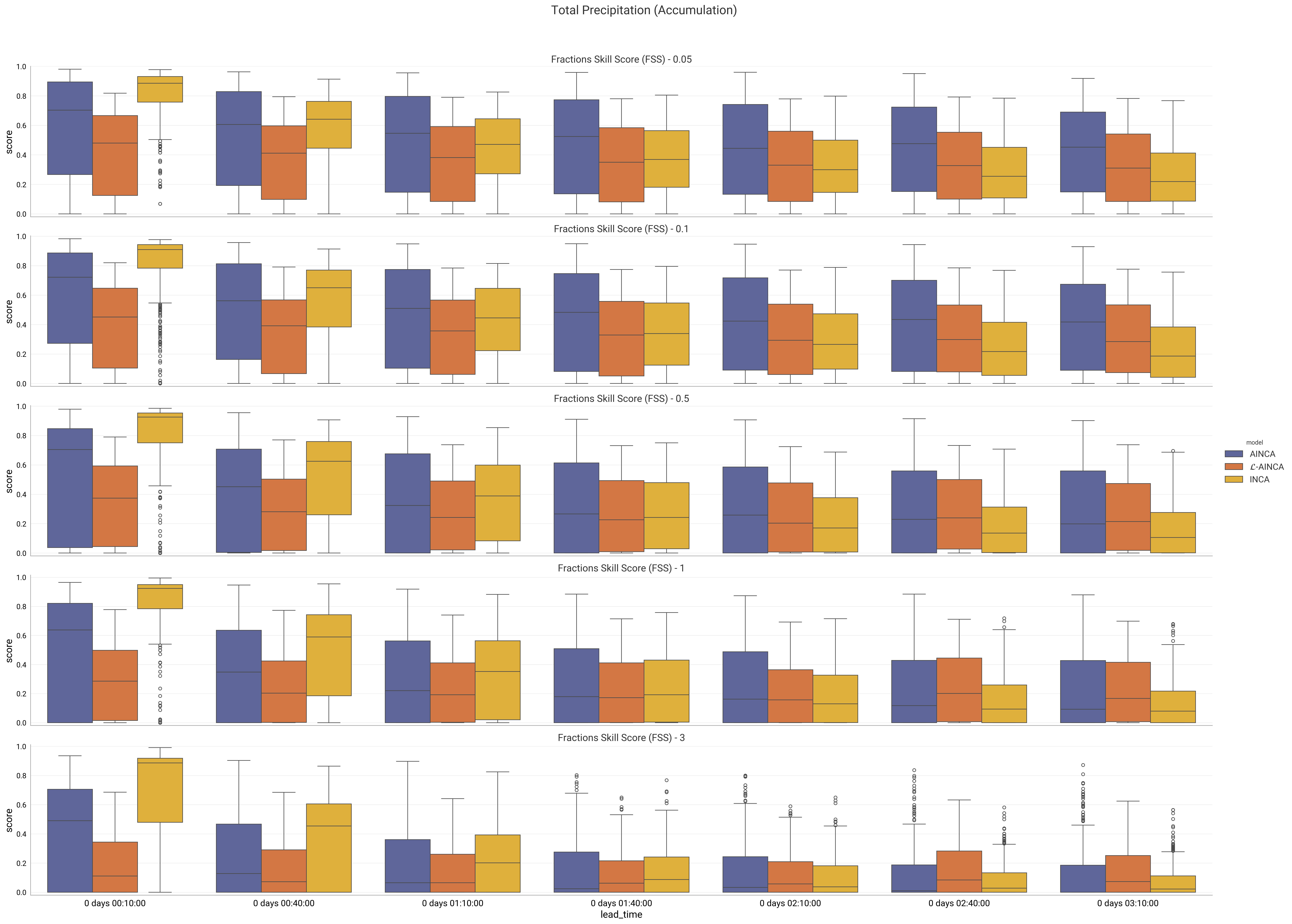}
    \caption{Boxplot time series of the fraction skill score distribution over the test set for precipitation events exceeding different thresholds (rows). Test aggregate.}
    \label{fig:boxplot_precip}
\end{figure}
\clearpage
\bibliographystyle{ametsocV6}
\bibliography{bibli}

\end{document}